\documentclass[preprint,superscriptaddress]{revtex4-1}

\usepackage{graphicx}
\usepackage{epstopdf}
\usepackage[ansinew]{inputenc}
\usepackage{array}
\usepackage{color}
\usepackage{amsmath}
\usepackage{amsxtra}
\usepackage{amstext}
\usepackage{amssymb}
\usepackage{latexsym}
\usepackage{float}
\usepackage{bm}
\usepackage{dsfont}

\usepackage{textcomp}
\usepackage[T1]{fontenc}

\begin{document}

\title{Pressure-induced commensurate stacking of graphene on boron nitride}
\author{Matthew Yankowitz}
\altaffiliation{Present Address: Department of Physics, Columbia University, New York, NY 10027, USA}
\affiliation{Physics Department, University of Arizona, Tucson, AZ 85721, USA}
\author{K. Watanabe}
\author{T. Taniguchi}
\affiliation{National Institute for Materials Science, 1-1 Namiki, Tsukuba 305-0044, Japan}
\author{Pablo San-Jose}
\affiliation{Instituto de Ciencia de Materiales de Madrid (ICMM-CSIC), Cantoblanco, 28049 Madrid, Spain}
\author{Brian J. LeRoy}
\email{leroy@physics.arizona.edu}
\affiliation{Physics Department, University of Arizona, Tucson, AZ 85721, USA}
\date{\today}

\begin{abstract}

Combining atomically-thin van der Waals materials into heterostructures provides a powerful path towards the creation of designer electronic devices. The interaction strength between neighboring layers, most easily controlled through their interlayer separation, can have significant influence on the electronic properties of these composite materials. Here, we demonstrate unprecedented control over interlayer interactions by locally modifying the interlayer separation between graphene and boron nitride, which we achieve by applying pressure with a scanning tunneling microscopy tip. For the special case of aligned or nearly-aligned graphene on boron nitride, the graphene lattice can stretch and compress locally to compensate for the slight lattice mismatch between the two materials. We find that modifying the interlayer separation directly tunes the lattice strain and induces commensurate stacking underneath the tip. Our results motivate future studies tailoring the electronic properties of van der Waals heterostructures by controlling the interlayer separation of the entire device using hydrostatic pressure.

\end{abstract}

\maketitle

\section*{Introduction}

The electronic properties of heterostructures of van der Waals (vdW) materials are expected to depend on the exact nature of the interactions between the composite layers. Previous work has focused on controlling the properties of these systems through the choice and ordering of the materials in the heterostructure, as well as the rotational alignment between layers~\cite{Geim2013}, but little has been done to explore the inerlayer separation degree of freedom. In bilayer graphene, for example, the electronic coupling between the two layers depends exponentially on their separation~\cite{Laissardiere2010}, controlling the effective mass of the charge carriers and the magnitude of the field-tunable band gap~\cite{McCann2013}. For graphene on atomically-heavy materials, such as WSe$_2$ or topological insulators, the strong substrate spin-orbit interaction (SOI) is predicted to strongly enhance the SOI in the graphene and possibly induce topologically non-trivial insulating states~\cite{Zhang2014,Wang2015}. The predicted magnitude of the SOI in the graphene also depends critically on the interlayer separation in such structures. Less immediately apparent, modifying the interlayer separation through pressure can also induce a commensurate match between two crystals with slight lattice mismatch at equilibrium. 

\begin{figure}
\newpage
\centering
\includegraphics[width=8.6cm]{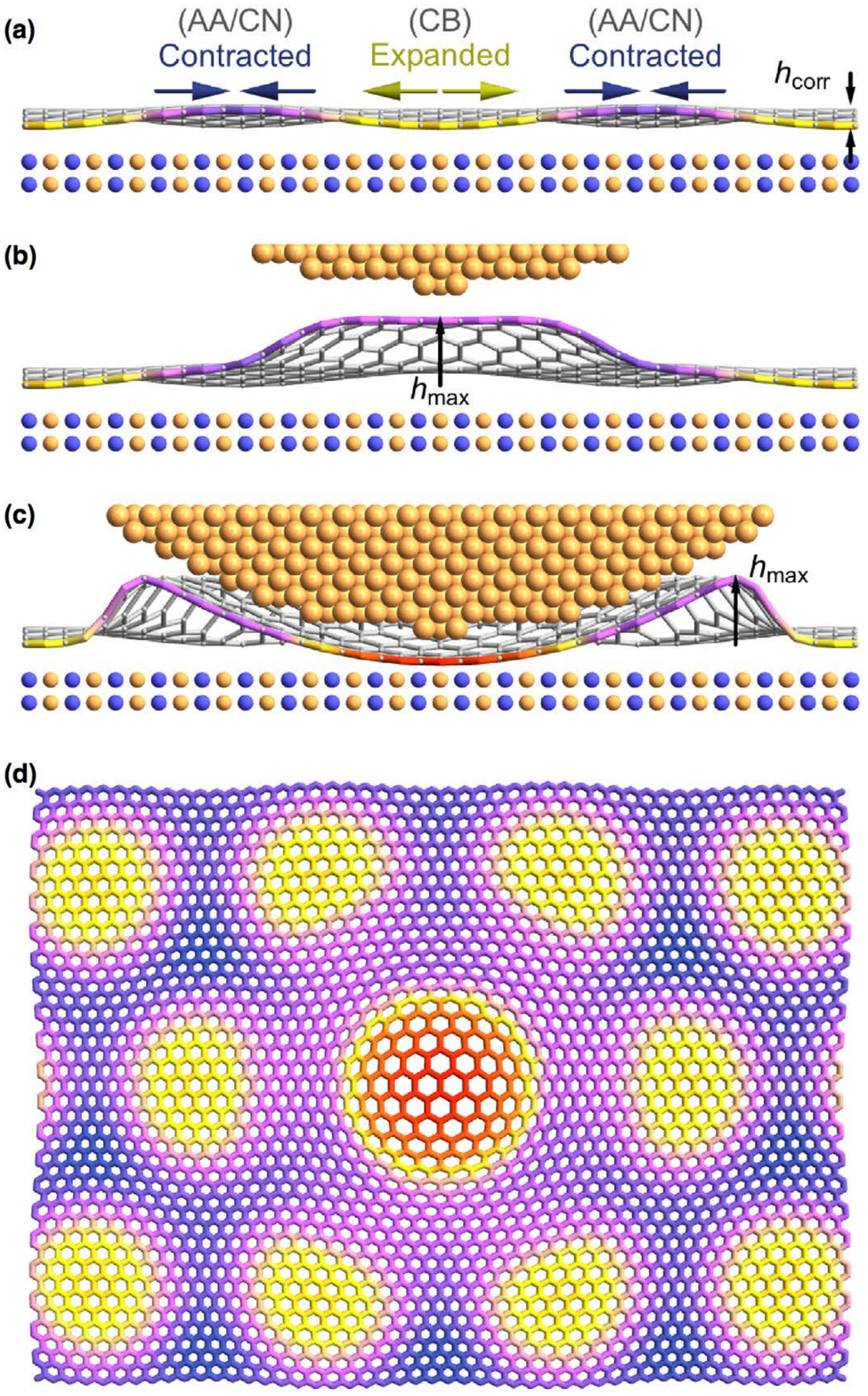} 
\caption{Schematic of graphene on hBN and the influence of an STM tip. (a) Schematic of an aligned graphene on hBN heterostructure. Due to the spatially modulated vdW adhesion potential, the graphene lattice periodically expands and contracts in-plane. An out-of-plane corrugation profile also develops, both matching the moir\'e. (b) In the presence of an STM tip, a vdW adhesion between the tip and graphene lifts the graphene off the surface of the hBN, modifying the strain field. (c) For an STM tip very close to the surface, the graphene is pushed closer to the hBN, enhancing the difference in the adhesion potential for different stacking configurations. The graphene lattice then expands to match the slightly longer lattice constant of the hBN. (d) Top view of (c), where the STM tip sits in the center of a moir\'e period (i.e. over a CB stacking configuration). The graphene lattice expands locally (red) to match the hBN lattice. Both the lattice constant and the spatial deformation have been scaled up for better visibility.}
\label{fig:schematic}
\end{figure}

Graphene on hexagonal boron nitride (hBN) is an excellent testbed for this effect, as a long-wavelength periodic interaction emerges when the two crystals are in near-rotational alignment due to their small lattice mismatch ($\delta \sim1.8$\%)~\cite{Giovannetti2007,Xue2011,Decker2011}. This moir\'e pattern spatially modulates both the electronic coupling and the van der Waals adhesion between the graphene and hBN lattices. The periodic modulation of the electronic potential leads to secondary Dirac cones in the graphene spectrum~\cite{Yankowitz2012}, while the modulation of the adhesion potential is expected to produce periodic in-plane strains of the graphene lattice. The latter arise because the adhesion potential is stronger for carbon-boron (CB) stacking than for any other lattice alignment. As a result, the graphene lattice expands locally around CB-stacked regions to increase the area of this favored stacking. This occurs at the expense of other stacking configurations, so that the total adhesion plus elastic energies is minimized~\cite{SanJose2014a}. A small out-of-plane lattice corrugation matching the moir\'e also develops to minimize the total potential energy of the system~\cite{Wijk2014,Jung2015,Kumar2015} (see Supplementary Note 4 and Supplementary Figure 7). Small electronic band gaps are expected to emerge for such a scenario, as the sublattice symmetry of the graphene is slightly broken due to the in-plane strain field~\cite{SanJose2014a,Bokdam2014,SanJose2014b,Jung2015}. A large enough enhancement of the adhesion modulation should cause the graphene to snap into a globally commensurate CB-stacked phase (i.e. graphene stretching uniformly to compensate for the lattice mismatch with hBN). The resulting heterostructure is expected to become a very high-mobility semiconductor with a sizable ($\sim50-200$ meV) band gap~\cite{Giovannetti2007,Bokdam2014}. Importantly, the strength of the adhesion modulation is controlled directly by the interlayer separation.

Here we demonstrate a path towards achieving control over this degree of freedom by demonstrating that pressure exerted by a scanning tunneling microscopy (STM) tip ~\cite{Mashov2010,Klimov2012,Xu2012,Altenburg2014,Meza2015} is capable of compressing or relaxing the interlayer separation locally between graphene and hBN. We also show that by modulating the interlayer separation we can control the degree of local commensurate stacking and the in-plane strain of graphene. This technique provides unprecedented control over the crystal structure of a 2D vdW heterostructure. 

\section*{Results}

\subsection*{Lifting graphene with an STM tip}

We first present evidence of the out-of-plane movement of the graphene lattice produced by the tip, depicted schematically in Figs.~\ref{fig:schematic}(b) and (c). We monitor the tunnel current $I$ as a function of the relative tip-sample separation $\Delta z$. The tunneling current is expected to scale exponentially with $\Delta z$ as 
\begin{equation} \label{tunnelcurrent}
I \propto e^{-\Delta z\sqrt{\frac{8m\phi}{\hbar^2}}},
\end{equation}
where $m$ is the electron mass and $\phi$ is the tunnel barrier height. This exponential approximation holds well for graphene on SiO$_2$, but fails for graphene on hBN (Fig.~\ref{fig:adhesion}(a)), independent of relative rotation angle (see Supplementary Note 2 and Supplementary Figure 3). In the latter case, $I(\Delta z)$ becomes strongly dependent on the specific tunneling parameters, with the tunnel current decay growing slower as the tip distance is brought closer to the surface. Furthermore, the decay is initially quadratic rather than exponential. Fig.~\ref{fig:adhesion}(b) shows a similar measurement with the tunnel current plotted on a logarithmic scale, further highlighting the initial regime of non-exponential decay. The departure from Eq.~\ref{tunnelcurrent} implies that the graphene is moving with the STM tip as it retracts from the sample, owing to a vdW attraction between the graphene and the tip. The vdW adhesion is apparently stronger between the tip and graphene than between the graphene and hBN, as evidenced by a visible $I(\Delta z)$ hysteresis between tip approach and tip retraction (see Supplementary Figure 3). This allows the tip to modify the interlayer separation (while conversely, the graphene is more strongly adhered to the SiO$_2$ substrate and is relatively immobile).

\begin{figure}
\newpage
\centering
\includegraphics[width=8.6cm]{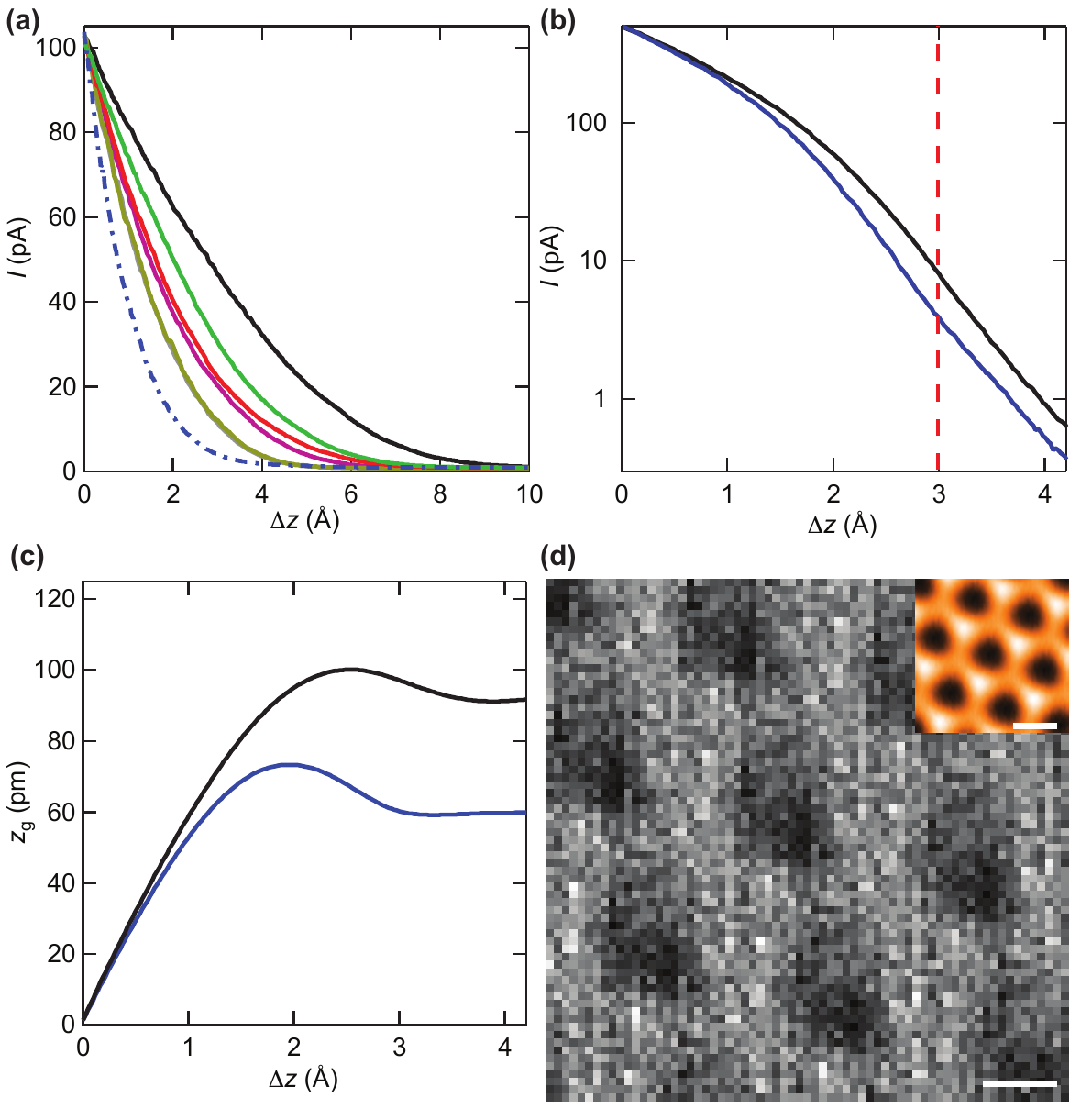} 
\caption{Tunneling current as a function of tip-sample separation. (a) Measurement of the tunnel current $I$ versus tip retraction distance $\Delta z$ for nearly-aligned graphene on hBN, starting with the tip in close proximity to the sample. The dot-dashed blue curve is taken on graphene on SiO$_2$ for reference, and exhibits the anticipated exponential decay. The remaining curves, from gold to black, represent decreasing sample bias (i.e. moving the tip closer to the surface), from 1 V to 0.05 V. The decay is initially parabolic, and the crossover point to exponential decay grows to larger $\Delta z$ as the sample bias is lowered. (b) Similar decay measurement plotted on a log scale on a CB (blue) and CN/AA (black) region. The transition from parabolic to linear occurs at $\Delta z$ of about 2 \AA. (c) Out-of-plane graphene movement relative to the hBN ($z_\mathrm{g}$) as a function of tip separation $\Delta z$. As the tip is initially retracted, the graphene moves with it, lifting away from the hBN. At just over 2 \AA, a maximum pulling distance is reached, and upon further tip retraction the graphene slowly relaxes back towards the hBN. (d) Spatial map of the tunneling current (dark is low and bright is high). The data is taken from the same set as in (b), at $\Delta z$ = 3 \AA. The inset displays the simultaneously acquired topography. The tunneling current is smaller in the moir\'e centers than along the boundaries, suggesting a spatial modulation in the ability of the tip to pull the graphene off the hBN substrate. The maps have been spatially averaged (see main text). Note that a similar pattern is exhibited at all $\Delta z$, as the blue curve is always below the black in (b). The scale bar is 5 nm for the main panel and 10 nm for the inset.}
\label{fig:adhesion}
\end{figure}

To account for the additional out-of-plane movement of the graphene sheet, we substitute $\Delta z$ in Eq.~\ref{tunnelcurrent} with $\Delta z-z_\mathrm{g}(z)$, where $z_\mathrm{g}(z)$ represents the movement of the graphene relative to the hBN substrate as a function of the tip position $z$. We plot the relative movement of the graphene in Fig.~\ref{fig:adhesion}(c), assuming an effective barrier height $\phi$ = 4 eV, as extracted from measurements acquired at large tip-sample separations. The tip initially lifts the graphene away from the hBN as it retracts. After around 2 \AA~ of retraction, the tip is no longer able to continue pulling the graphene, which then begins to slowly relax back towards the hBN substrate, as it is still under the influence of a vdW force from the tip~\cite{Altenburg2014}. It is important to note that the graphene is initially pushed towards the hBN by the tip, so the equilibrium separation lies somewhere at $z_\mathrm{g} > 0$. The blue and black curves in Figs.~\ref{fig:adhesion}(b) and (c) are taken in the center and along the  boundaries of the moir\'e, respectively, and exhibit a spatial variation in the maximum pulling amplitude of the tip. The variations can be further highlighted by plotting a spatial map of the tunneling current at a fixed tip retraction distance $\Delta z$, as in Fig.~\ref{fig:adhesion}(d). The spatial variation in the current matches the topographic moir\'e pattern, suggesting modulations in the magnitude of the out-of-plane graphene pulling by the tip due to the underlying spatial modulations in the adhesion potential between the graphene and the hBN. 

\begin{figure}
\newpage
\centering
\includegraphics[width=8.6cm]{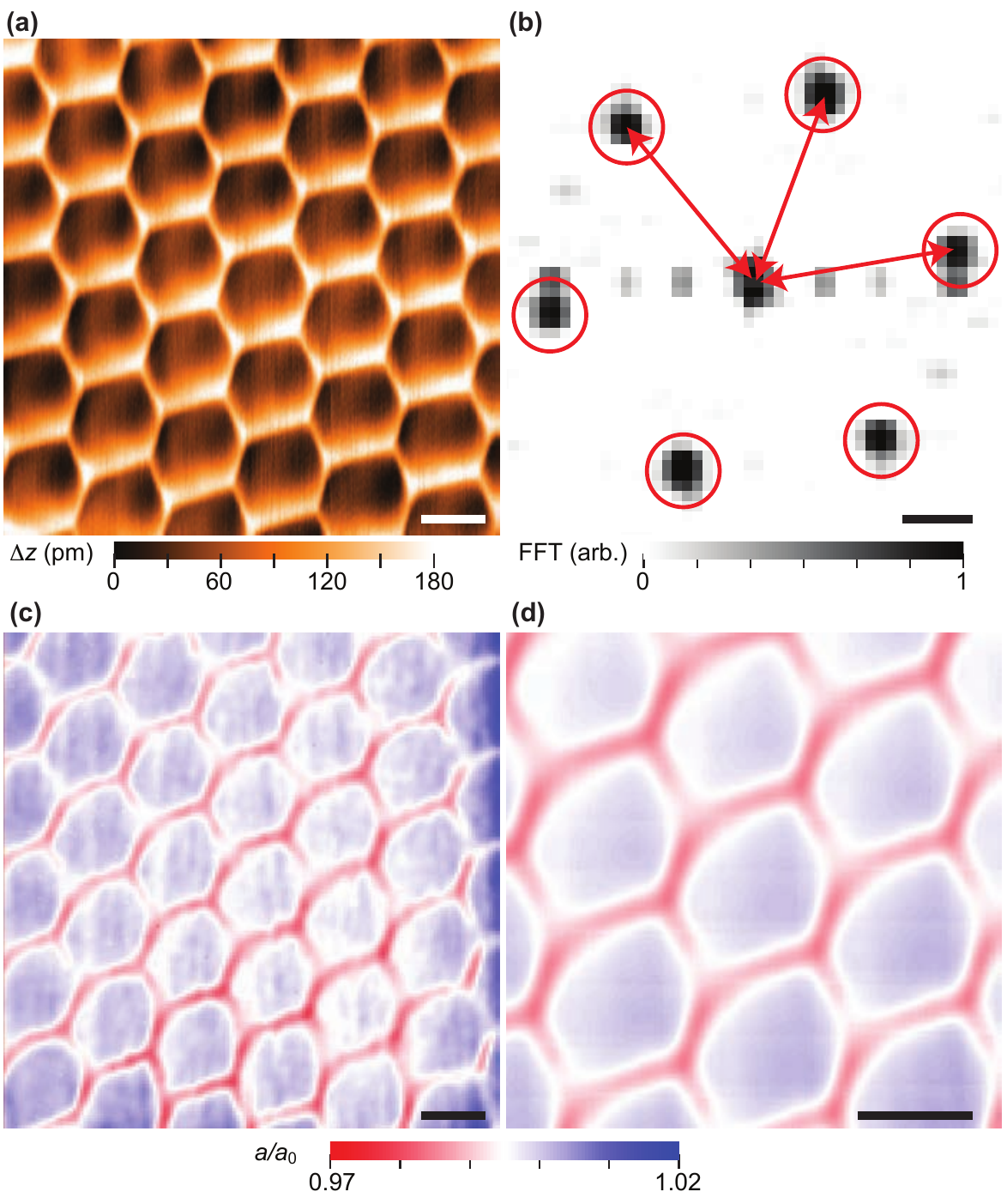} 
\caption{Method for generating strain maps. (a) Atomically-resolved topography of nearly-aligned graphene on hBN. The topography was acquired with a sample voltage of $V_\mathrm{s}$ = 0.3 V and a tunneling current of $I_\mathrm{t}$ = 200 pA. (b) Fourier transform of a 4 nm x 4 nm region of (a), showing six resonances representing the hexagonal graphene lattice (red circles). The red arrows depict the measurement of the lattice constant in each direction. (c) Plot of the average length of the three lattice directions, as measured in (b) for each point in the topographic map. The points are normalized by the equilibrium graphene lattice constant $a_0$. (d) Spatially-averaged strain map, generated by averaging (c) over a few moir\'e unit cells. The scale bars are 10 nm for (a), (c) and (d), and 10 nm$^{-1}$ for (b).}
\label{fig:method}
\end{figure}

\subsection*{Modifying commensuration with interlayer spacing}

The relative adhesion potentials between the CB, CN (carbon-nitrogen), and AA (hexagons atop one another) stacking configurations depend on the interlayer separation between the two materials (see Supplementary Note 4 and Supplementary Figure 6). To understand how the in-plane strains in the graphene lattice depend on the interlayer separation, and to show how they can be controlled through tip pressure, we have acquired atomically resolved topographic maps of nearly-aligned graphene on hBN heterostructures (Fig.~\ref{fig:method}(a)) with varying tunnel resistance (which controls tip-sample separation and therefore the interlayer separation). All measurements were performed in ultra-high vacuum at a temperature of 4.5 K. From a topographic map, we take small (4 nm x 4 nm) areas, perform a Fourier transform (Fig.~\ref{fig:method}(b)), and extract the average length $a$ of the three resonances due to the hexagonal graphene lattice. We then create a map of the average graphene lattice constant normalized by the equilibrium length ($a/a_0$, with $a_0$ = 2.46 \AA) as a function of position (Fig.~\ref{fig:method}(c)). Finally, to enhance the clarity of these strain images we average each point in the moir\'e unit cell with all other equivalent sites in the strain image (Fig.~\ref{fig:method}(d)).

Figs.~\ref{fig:strain}(a)-(c) show spatially-averaged STM topography images taken over the same area of a nearly-aligned graphene on hBN sample with decreasing tip-sample separation. The hexagonal stacking boundaries in the measured moir\'e pattern grow sharper as the tip moves closer to the surface, exerting an increasing pressure. Below a critical tip separation, the stacking boundaries appear atomically and sub-atomically sharp, and a hysteresis eventually develops in their positions between the forward and backward scan directions (Fig.~\ref{fig:strain}(c) and Supplementary Figure 1). This observation clearly points to a strong influence of the tip on the graphene lattice. If the sample were unperturbed by the tip, the appearance of the topography, and in particular the measured thickness of the stacking boundaries would correspond to the equilibrium sample configuration, and should not depend on the tip pressure except for a local density of states (LDOS) component which can be eliminated (see Supplementary Note 1). The graphene lattice strain maps for the different characteristic profiles of the moir\'e topography are shown in Figs.~\ref{fig:strain}(d)-(f). Like the topography, these are not equilibrium strain fields but rather local strains under the tip that dynamically evolve during the scan in response to the moving tip interaction. We identify three typical and qualitatively different spatial patterns in this dynamical strain. Stacking boundaries can appear thick, but are expanded relative to the CB regions (large tip-sample separations, Fig.~\ref{fig:strain}(d)). This is opposite to the equilibrium expectation. Alternatively, boundaries can appear thin, and are compressed relative to CB regions (intermediate tip separations, Fig.~\ref{fig:strain}(e)). Finally, boundaries can exhibit hysteretic behavior and broken 3-fold symmetry, and the entire graphene lattice is expanded relative to equilibrium (smallest tip separations, Fig.~\ref{fig:strain}(f)). The response of the sample to the tip is so strong that, within the limits of our STM measurements, it is never possible to measure the equilibrium configuration of the heterostructure (i.e. even at very large tip-sample separations, the graphene is still lifted off the hBN). The apparently sharp boundaries in Fig.~\ref{fig:strain}(c) in particular, also observed in our previous work~\cite{Yankowitz2014}, are therefore not an equilibrium configuration.

Interestingly, we observe qualitatively similar behavior in slightly misaligned samples as well. Specifically, we observe the three different strain profiles as a function of tip-sample separation in all moir\'e areas studied with periods varying from 14 nm (essentially perfect alignment) down to about 6 nm (below which the behavior may persist, but our analysis is no longer sensitive as the size of our Fourier transform window becomes comparable to the entire moir\'e unit cell). As an example, Supplementary Figure 2 shows strain maps for an 8 nm moir\'e period. This observation is in stark contrast to the results of Ref.~\cite{Woods2014}, the reasons for which will be discussed in our model below and in Supplementary Note 6.

\begin{figure}
\newpage
\centering
\includegraphics[width=8.6cm]{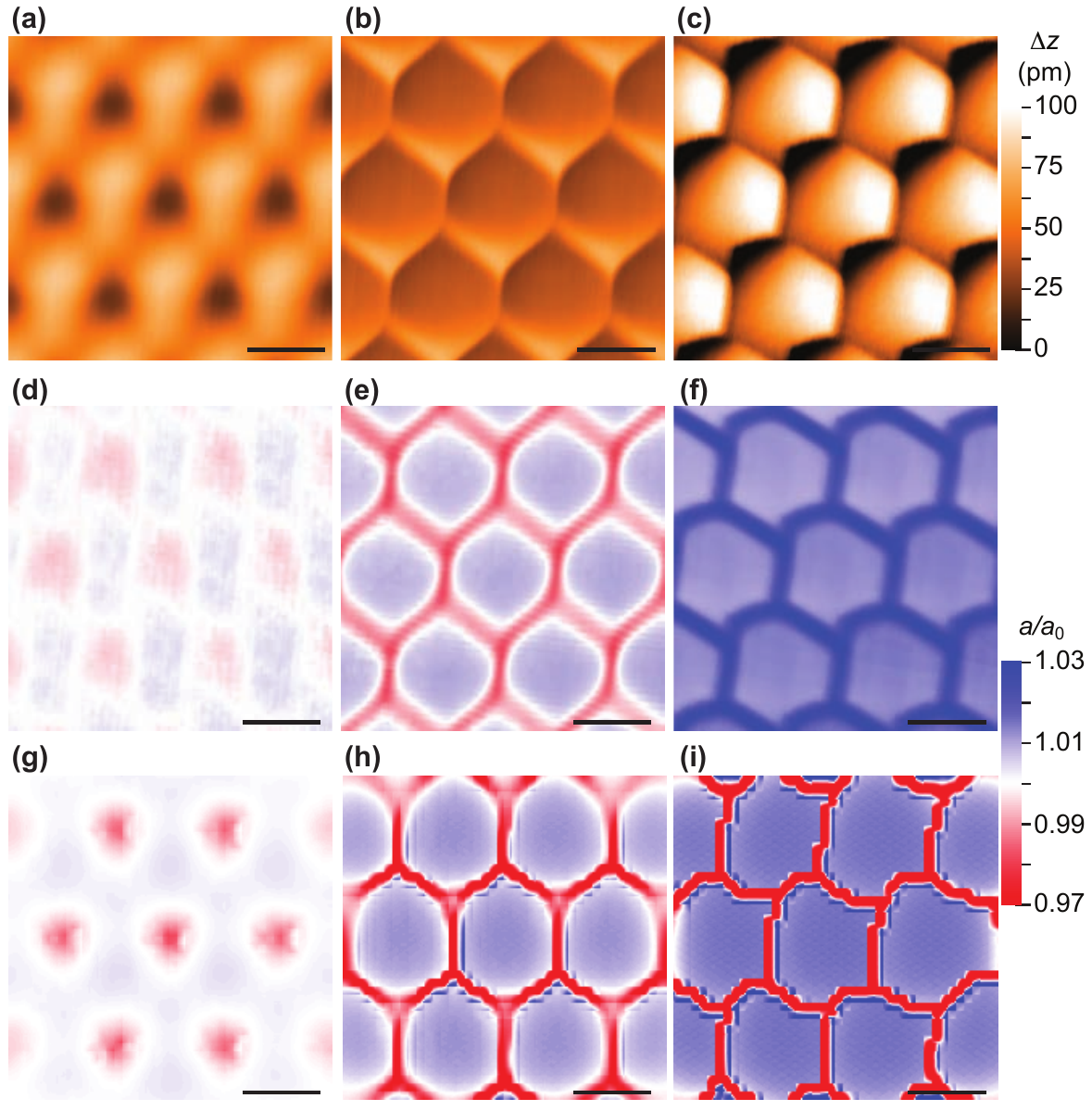} 
\caption{Topography and strain maps in different interaction regimes. (a-c) Spatially-averaged topography maps acquired over the same region of a nearly-aligned sample, with the tip moving progressively closer to the sample. The appearance of the stacking boundaries becomes sharper as the tip moves closer, and becomes hysteretic and asymmetric in (c). (d-f) Experimental spatially-averaged strain maps generated from the topographic maps of (a)-(c). In (d) the graphene lattice is compressed in the moir\'e centers and expanded along the boundaries. The opposite behavior is observed in (e). In (f), the graphene lattice constant for the entire map is expanded, as the system is in a strongly interacting, hysteretic regime. (g-i) Simulated strain maps, showing excellent agreement with the experimental results. The disagreement in the dynamical strain at the stacking boundaries between (f) and (i) is attributed to the absence of out-of-plane buckling in the simulation. The tunneling parameters are (a) $V_\mathrm{s}$ = 0.5 V and $I_\mathrm{t}$ = 50 pA, (b) $V_\mathrm{s}$ = 0.5 V and $I_\mathrm{t}$ = 900 pA, and (c) $V_\mathrm{s}$ = 0.05 V and $I_\mathrm{t}$ = 100 pA. All scale bars are 10 nm.}
\label{fig:strain}
\end{figure}

\subsection*{Theoretical analysis}

We have simulated the dynamical strain of the graphene lattice under a scanning tip using a simple adhesion model between graphene and hBN (see Methods and Supplementary Note 4 for full details, as well as Supplementary Movies 1-4 for animations). In our model, the graphene sticks to a parabolic tip, and can thus be locally compressed against or separated away from the hBN substrate. Figs.~\ref{fig:strain}(g)-(i) show the strain maps obtained for decreasing tip-sample separations, which exhibit excellent agreement overall, both qualitatively and quantitatively with their experimental counterparts. The three characteristic spatial patterns arise naturally when the effective interaction between the tip and the equilibrium stacking boundaries changes with $z$ from attractive, to repulsive, and to strongly repulsive. In the attractive regime, the graphene under the tip is lifted off the hBN surface, lowering the adhesion potential. The stacking boundaries are then attracted to the scanning tip, and as a result the graphene lattice appears to be expanded along the stacking boundaries (Fig.~\ref{fig:strain}(g)). In the repulsive regime, the tip is pushing down on the sample, increasing the adhesion energy modulation. The CB-stacked regions then become expanded under the tip, up to the maximum static value $a/a_0 =1+ \delta$ (local commensurate stacking) at high pressure, and the stacking boundaries are pushed away (see Fig.~\ref{fig:schematic}(d) for a schematic of the graphene lattice strain when the tip sits above the CB center of the moir\'e). As the tip scans the sample, the commensurate area underneath (red in the schematic) moves with it, and the stacking boundaries are likewise pushed along (Fig.~\ref{fig:strain}(h)). If the tip pressure is strong enough, the stacking boundaries are pushed until, eventually, they irreversibly snap back under the tip (Fig.~\ref{fig:strain}(i)). This abrupt snapping results in the observed hysteretic behavior with tip scan direction, and a breaking of the characteristic 3-fold symmetry of the moir\'e pattern (note that the expanded hysteretic boundaries that develop in this regime may be explained by sudden out-of-plane delamination of graphene in front of the tip, a possibility not included in our model, see Supplementary Note 4). 

The notable success of our simulations in reproducing the experimental dynamical strain maps allows us to confidently remove the tip from the simulations, in order to understand the equilibrium configuration of the graphene lattice. We find that the observed phenomenology is consistent with intrinsic adhesion potential differences~\cite{Sachs2011,NeekAmal2014} of $V_\mathrm{AA}-V_\mathrm{CB}=16$ meV per graphene unit cell, similar to the values from \textit{ab initio} calculations~\cite{Wijk2014}. Importantly, our results are not consistent with an adhesion potential difference of zero (nor an infinitely stiff graphene lattice). The corresponding strain of the graphene at equilibrium (without a tip) is rather weak, and varies almost sinusoidally between $\pm 0.3 \%$ (see Supplementary Figure 10). This is in stark contrast to the dynamical strain maps, which may appear much sharper spatially and in excess of $\pm 1 \%$. These dynamical strain effects are important to consider in all scanning probe measurements of graphene on hBN~\cite{Yankowitz2012,Woods2014} (see Supplementary Note 6).

\section*{Discussion}

We have demonstrated unprecedented control of the atomic structure of graphene by locally modifying the interaction strength with an hBN substrate through pressure applied with an STM tip. This allowed us in particular to induce and directly image tunable in-plane strains and local commensurate stacking. While a globally commensurate graphene on hBN structure is expected to exhibit an electronic band gap, we do not observe any signatures of a gap in our tunneling spectroscopy measurements of the local density of states (see Supplementary Note 3) for any applied tip pressure. When the tip is far from the sample, such that it remains incommensurate, the tip likely screens the many-body interactions responsible for the development of the band gap typically observed in transport experiments~\cite{Bokdam2014,Song2013,Slotman2015}. When the graphene is commensurate with the hBN, the gap is expected to be of order 50 meV even before the consideration of potential many-body enhancement~\cite{Giovannetti2007}. Therefore, it may be seem surprising that we also do not observe a band gap in tunneling spectroscopy even in the case where the tip is very close to the sample, such that the graphene is commensurate with the hBN underneath the tip. However, the lack of observed band gap is a consequence of the local nature of the applied pressure in our experimental setup. A gap of magnitude $\Delta$ corresponds to the localization of states of typical wavelength $\lambda_\Delta=h v_\mathrm{F}/\frac{\Delta}{2}$. For the anticipated band gap $\Delta \approx$ 50 meV, states must be localized on length scales of order 100 nm. In our work, our model predicts that the area of the graphene forced into a commensurate state with the hBN is confined to approximately one moir\'e period, of order 10 nm (see Fig.~\ref{fig:schematic}(d)). Thus, the lack of a band gap in tunneling spectroscopy is to be expected because the commensurate area is considerably smaller than the requisite localization area (see Supplementary Notes 3 and 5 for further details about the tunneling spectroscopy measurements and their theoretical modeling).

This suggests a natural extension of our work, where a graphene sheet is forced into a commensurate state with hBN over the entire sample area. Fortunately, the technique of applying pressure to a vdW heterostructure is very easily generalizable to the scale of the entire device using hydrostatic or diamond anvil pressure cells. In graphene on hBN in particular, we anticipate a globally commensurate state to emerge under a hydrostatic pressure of roughly 150 MPa (see Supplementary Note 4 and Supplementary Figure 8), characterized by the absence of a moir\'e pattern and a large band gap due to globally broken sublattice symmetry in the graphene. More generally, global control of the interlayer separation through pressure in other vdW heterostructures should enable exciting new experimental designs and result in the emergence of many novel electronic device properties.

\section*{Methods}

\subsection*{Sample preparation and measurement details}

Chemical vapor deposition (CVD) grown graphene was transfered onto mechanically exfoliated hexagonal boron nitride resting on a Si/SiO$_2$ substrate. The devices were annealed at 350 $^{\circ}$C in a mixture of argon and hydrogen, then at 300 $^{\circ}$C in air. Similar results to those reported here were observed in preliminary work with exfoliated graphene flakes as well.

All the STM measurements were performed in ultra-high vacuum at a temperature of 4.5 K using a tungsten tip. The tunneling resistance was varied over five orders of magnitude by controlling the sample bias and tunneling current. We note that tip geometries are somewhat random between different tips, and between different tip shaping procedures on the same tip. Because the nature of the tip ending is also important for determining the interaction strength with the substrate, comparing tunneling resistances between different measurements is not itself a sufficient metric for determining the amount of compression or relaxation of the graphene relative to the hBN.

\subsection*{Tip preparation}

Tungsten tips were prepared by electrochemical etching, and further shaped \textit{in situ} when necessary by applying electrical pulses of 5 - 10 V on the Au contacts far from the graphene sample. The lattice deformation effects detailed here have been observed with every tip (tens of tips measured in total) and over tens of pulse cycles per tip. We note that qualitatively similar moir\'e scale lattice deformations have been observed in graphene on Ir(111) with AFM using a tip intentionally terminated with a carbon monoxide molecule~\cite{Boneschanscher2014}. While we cannot rule out that a deformable tip could have some influence on our results, we are confident that the primary source of the effects we present can be explained by our proposed model for a number of reasons. First, because we do not intentionally terminate our tips with a deformable molecule, it is very unlikely that we would observe similar results across all of our tips and pulse cycles if such a deformable tip ending were being randomly picked up every time. Second, the deformable tip ending would have to be metallic to be relevant for our tunneling measurements. While our samples may have water, hydrogen, or other small molecule adsorbates, they should certainly be free of metallic contaminants to unintentionally attach to the end of every tip. Further, we observe our reported behavior even with brand new tips which are landed directly onto the graphene. Third, we observe sub-atomically sharp discontinuities in the topography only on the moir\'e length scale (in contrast to previous reports showing such behavior on the atomic scale using a cobalt atom dragged across the surface of the sample~\cite{Stroscio2004}). No similar model can easily explain our observation of smooth atoms except at moir\'e boundaries in the hysteretic regime, which would require a much longer deformation length scale and a strong preference for irreversible topographic discontinuities only at special sites on the moir\'e. This suggests the discontinuities instead arise from lattice deformations in the graphene at moir\'e boundaries as we argue in our model. Finally, we observe a saturation of the graphene lattice constant expansion at just under 2\% in the hysteretic regime (excluding the boundaries which exhibit irreversible discontinuities), consistent with a commensurate structural transition (as this is roughly the lattice mismatch between graphene and hBN). We have never observed significantly larger lattice deformations. We would not anticipate such a bound if this effect were due to a deformable tip, providing further compelling evidence that the apparent lattice deformations we observe are primarily due to a modification of the graphene lattice itself, as proposed in our model.

\subsection*{Theoretical model}

An overview of our theoretical model is as follows (see the Supplementary Note 4 for full details). The STM tip is approximated by a paraboloid of radius $R$ around its apex, hovering at height $h_0$ relative to a relaxed reference plane (taken as the graphene position at the CB-stacked regions -- recall that graphene is slightly corrugated due to non-uniform adhesion to hBN). We assume that the vertical graphene displacement conforms to the tip profile as long as it does not exceed a certain height, $h_\mathrm{max}$, see Fig. \ref{fig:schematic}c. Otherwise graphene takes on the equilibrium vertical displacements at each stacking. We assume a certain in-plane distortion $\bm{u}(\bm{r})$ of the sample, relative to the relaxed moir\'e pattern, which we want to determine. We construct a smooth interpolation of the ab-initio adhesion potentials $V_\mathrm{S}(z)$ between different graphene/hBN stackings, where $z$ is the separation between the two crystals. Using the interpolated potential, we evaluate the total adhesion energy per unit area for a given field $\bm{u}(\bm{r})$. At each $\bm{r}$, the value of $z$ is constrained by the tip profile, as described above. To this adhesion energy, we add the corresponding elastic energy associated to $\bm{u}(\bm{r})$. We discretize $\bm{r}$, and express the total energy as a function of the finite set of $\bm{u}$ on the discrete mesh. We minimize the total energy, using conjugate gradient methods, and find the deformation $\bm{u}(\bm{r})$ at equilibrium. We then obtain the dynamical strain as measured by the tip by performing this sample relaxation as the tip moves across the sample at a constant height $h_0$. The model has no unconstrained free parameters, as all can be roughly estimated experimentally.

\subsection*{Data availability}
The data that support the findings of this study are available from the corresponding author upon request.

\bibliography{YankowitzManuscript}
\bibliographystyle{apsrev4-1}

\newpage

\section*{Acknowledgements}
We thank J. Sanchez-Yamagishi and P. Jarillo-Herrero for a device measured during the early development stages of this work. The work at Arizona was partially supported by the U. S. Army Research Laboratory and the U. S. Army Research Office under contract/grant number W911NF-14-1-0653 and the National Science Foundation DMR-0953784. P. S-J. was supported by the Spanish Ministry of Economy and Innovation through Grant No. FIS2011-23713 and the Ram\'on y Cajal programme.

\section*{Author contributions}
M.Y. and B.J.L. designed the experiments.  M.Y. fabricated the graphene on hBN devices and performed the STM experiments. K.W. and T.T. provided the single crystal hBN. P.S.-J. performed the theoretical calculations. All authors participated in the data discussion and writing of the manuscript.

\section*{Competing financial interests}
The authors declare no competing financial interests.

\newpage

\newcommand{\mat}[1]{\hat{#1}}
\newcommand{\vect}[1]{\bm{\mathrm{#1}}}
\renewcommand{\figurename}{Supplementary Figure}
\renewcommand{\thefigure}{~\arabic{figure}}
\renewcommand{\tablename}{Supplementary Table}
\renewcommand{\thetable}{~\arabic{table}}

\section*{Supplementary Figures}

\begin{figure}[h]
\centering
\includegraphics[width=8.6cm]{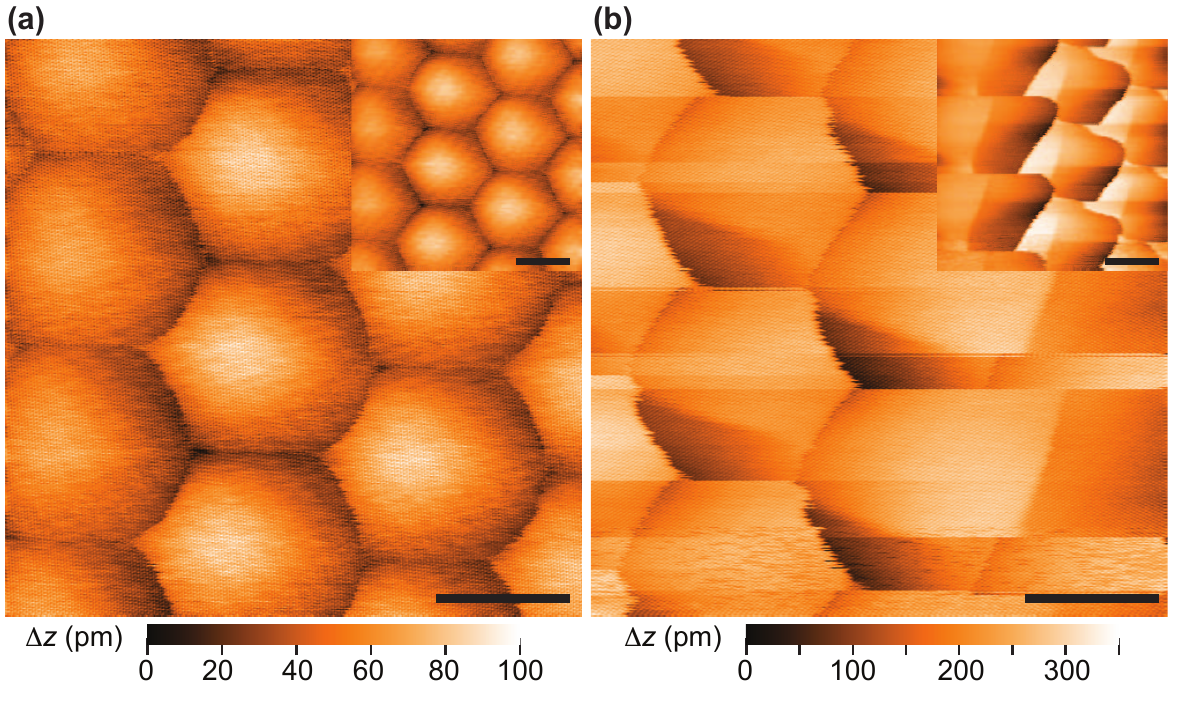} 
\caption{Topography in the hysteric regime. The tip gets closer to the sample from (a) to (b). The tip scans from left to right in both. The respective insets show the opposite scanning direction. The walls are pushed in the scanning direction of the tip. The walls are (a) atomically and (b) sub-atomically sharp, indicating a non-equilibrium measurement of the graphene, and are pushed further as the tip is in a more invasive regime. The tunneling parameters are (a) $V_\mathrm{s}$ = 0.3 V and $I_\mathrm{t}$ = 150 pA and (b) $V_\mathrm{s}$ = 0.025 V and $I_\mathrm{t}$ = 150 pA. All scale bars are 10 nm.}
\label{fig:topography}
\end{figure}

\newpage

\begin{figure}[t]
\centering
\includegraphics[width=8.6cm]{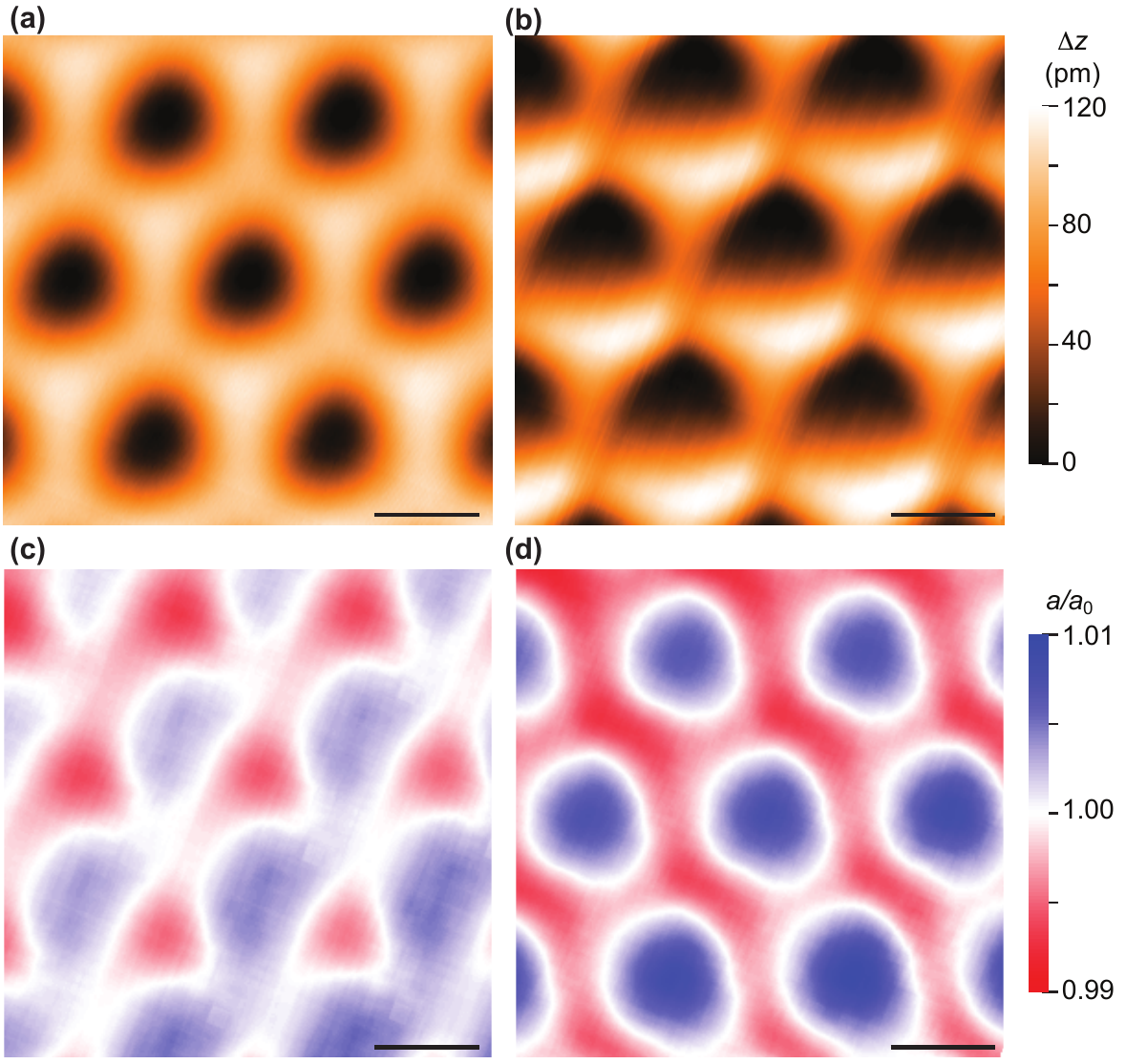} 
\caption{Topography and strain maps in different interaction regimes. (a)-(b) Spatially-averaged topography maps acquired over the same region of a slightly-misaligned sample (8 nm moir\'e), with the tip moving progressively closer to the sample. (c)-(d) Spatially-averaged strain maps generated from the topographic maps of (a)-(b). In (c) the graphene lattice is compressed in the moir\'e centers and expanded along the boundaries. The opposite behavior is observed in (d). These behaviors are consistent with the nearly-aligned samples. The tunneling parameters are (a) $V_\mathrm{s}$ = 0.95 V and $I_\mathrm{t}$ = 100 pA and (b) $V_\mathrm{s}$ = -0.15 V and $I_\mathrm{t}$ = 100 pA. All scale bars are 5 nm.}
\label{fig:strain}
\end{figure}

\newpage
~
\newpage

\begin{figure}[t]
\centering
\includegraphics[width=8.6cm]{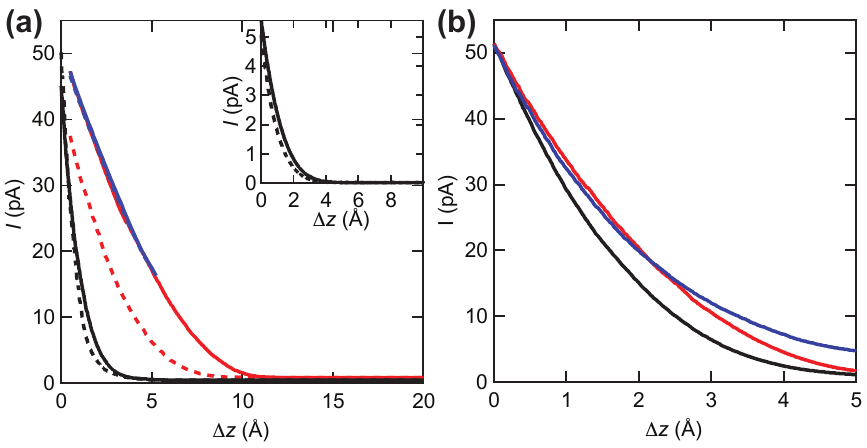} 
\caption{Decay of the tunneling current. (a) Tunneling current versus $\Delta z$ for both the tip retracting (solid) and approaching (dashed) the sample. The black curves are acquired on graphene on SiO$_2$, exhibiting roughly exponential decay and virtually no hysteresis. The red curve is acquired on graphene on hBN, showing significant departure from exponential decay and large hysteresis. The blue curve is acquired similarly, but with only a small retraction such that the tunneling current is always finite. In this case, no hysteresis is observed. (Inset) Graphene on hBN with a very large initial tip-sample separation. In this case, the decay is nearly exponential and non-hysteric. (b) Decay curves for misaligned graphene on hBN for decreasing initial tip-sample separations (black to red to blue). Similar behavior to the nearly-aligned case is observed, marked by significant departure from exponential decay.}
\label{fig:tunnelingcurrent}
\end{figure}

\newpage
~
\newpage

\begin{figure}[t]
\centering
\includegraphics[width=8.6cm]{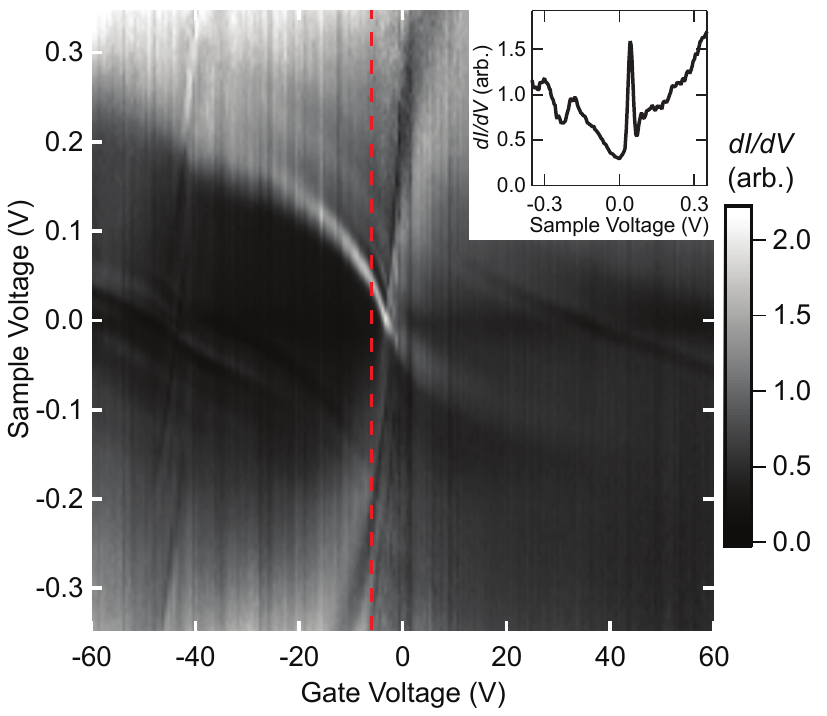} 
\caption{Gate map of nearly-aligned graphene on hBN acquired in a strongly invasive regime. The main Dirac point is prominently visible, though there is no obvious associated band gap. Replica Dirac point features are also observed in the conduction and valence bands. (Inset) Cut of the dI/dV taken at V$_\mathrm{g}$ = -5 V (the charge neutrality point). The density of states at the Dirac point remains finite, implying the lack of a band gap.}
\label{fig:gatemap}
\end{figure}

\newpage
~
\newpage

\begin{figure}[t]
\centering
\includegraphics[width=0.45\columnwidth]{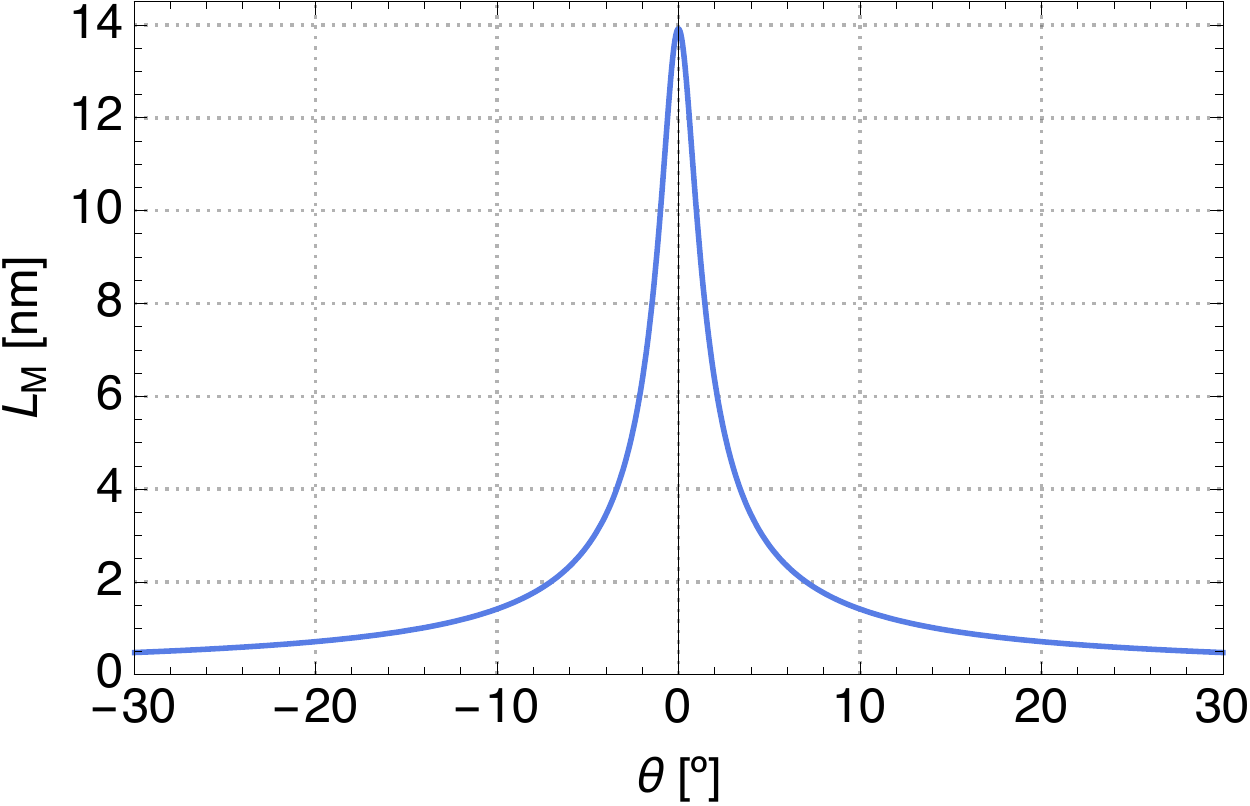} 
\caption{The moir\'e period $L_\mathrm{M}=|\mat{A}_\mathrm{i}|$ as a function of twist angle $\theta$, for mismatch $\delta=1.8\%$.}
\label{fig:LM}
\end{figure}

\newpage
~
\newpage

\begin{figure}[t]
\centering
\includegraphics[width=0.35\columnwidth]{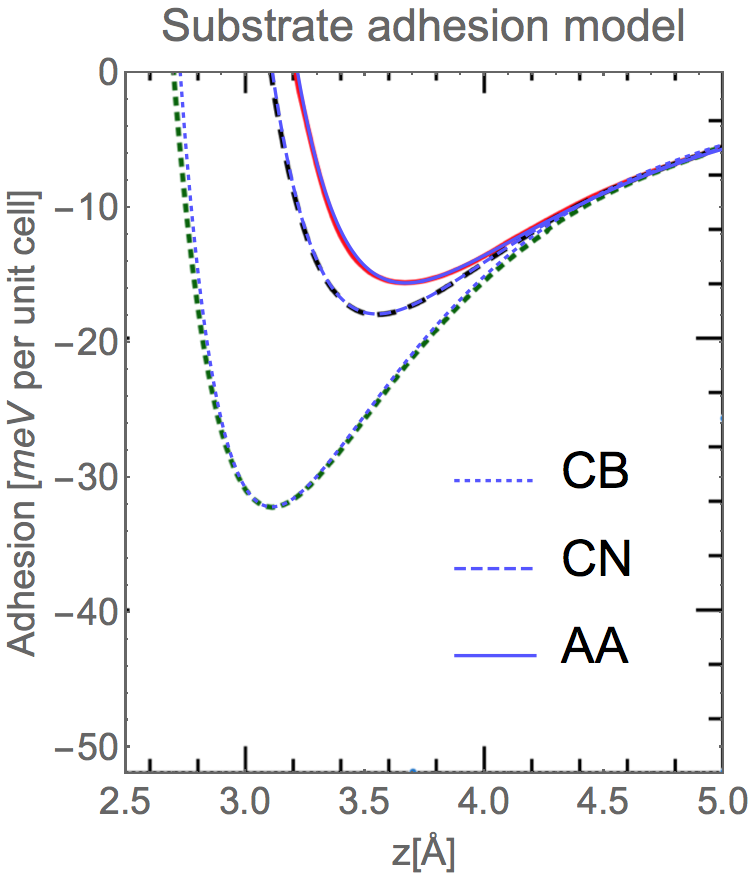} 
\caption{Theoretical adhesion energy versus stacking. Fit (in blue) to microscopic results (Ref.~\cite{Wijk2014}, in red, black and green, reproduced here with permission) for the adhesion energy $V_\alpha(z)$ at different stackings $\alpha=AA, CN, CB$.}
\label{fig:fit}
\end{figure}

\newpage
~
\newpage

\begin{figure}[t] 
\centering
\includegraphics[width=0.9 \textwidth]{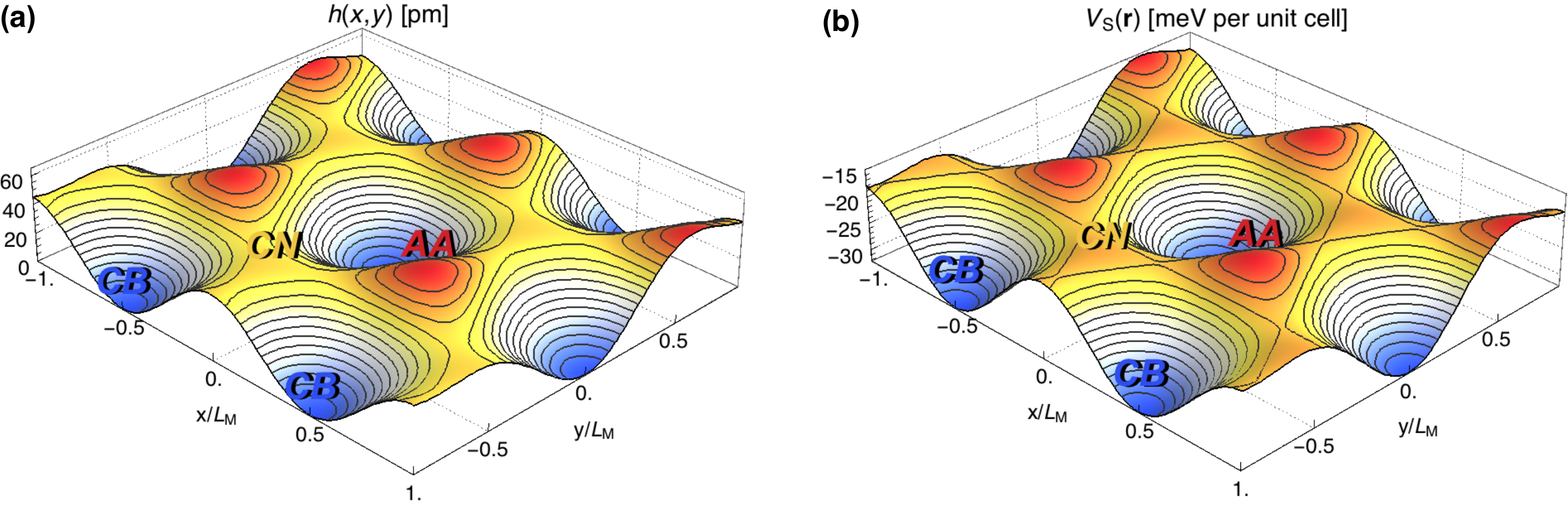} 
\caption{Height and potential landscape on the moir\'e cell. (a) Substrate-induced corrugation $h(\vect{r})$ of the graphene sample at equilibrium, relative to the $z=d_\mathrm{CB}$ plane. (b) Adhesion energy to the substrate per graphene unit cell $V_\mathrm{S}(\vect{r})$.}
\label{fig:profiles}
\end{figure}

\newpage
~
\newpage

\begin{figure}[t] 
\centering
\includegraphics[width=0.99 \textwidth]{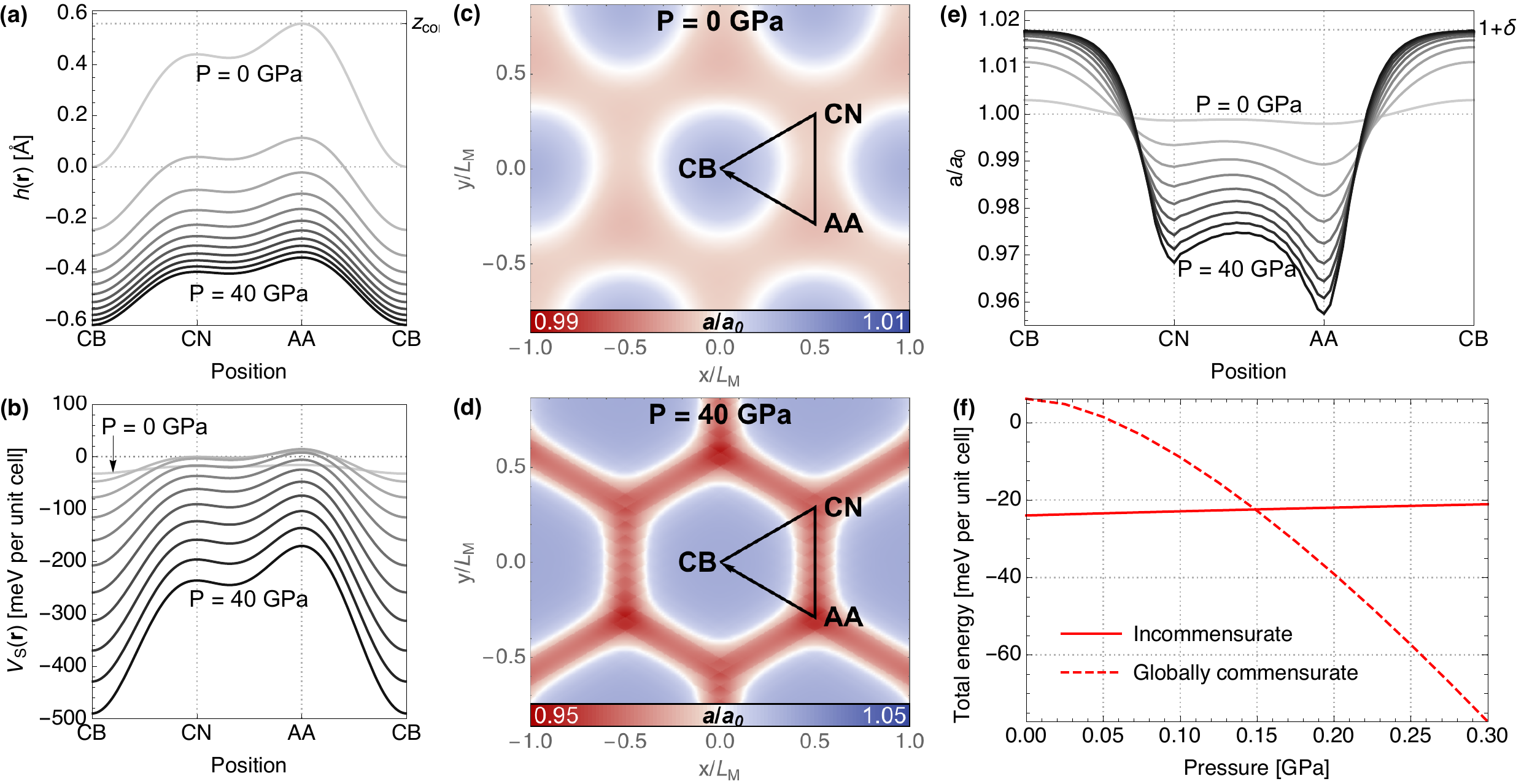} 
\caption{Graphene strain and energy versus hydrostatic pressure. Pressure dependence of (a) equilibrium corrugation $h(\vect{r})$ and (b) adhesion energy per unit cell, $V_\mathrm{S}(\vect{r})+V_\mathrm{P}(d_\mathrm{CB}+h(\vect{r}))$. Normalized average lattice constant $a/a_0=\frac{1}{2}\mathrm{Tr}\mat{u}+1$ at equilibrium for (c) zero hydrostatic pressure $P=0$, and (d) high pressure $P=40$ GPa. (e) Cut of local normalized lattice constant $a/a_0$ along the white path marked in (c,d), as pressure is increased. In all cases, the total area of the sample is forced to remain unchanged, thus precluding a transition to a globally conmensurate phase with uniform $a/a_0=1+\delta\approx 1.018$. (f) Total energy (elastic plus adhesion) per graphene unit cell at equilibrium versus pressure (solid line). The energy is measured relative to that of graphene in vacuum ($U=0$). The dashed line corresponds to the energy of the globally commensurate sample, with uniform strain $\delta$ and $h(\vect{r})=0$. Said configuration is preferred over the incommensurate moir\'e strain superlattice above a critical pressure $P_\mathrm{c}=0.15$ GPa.}
\label{fig:pressure}
\end{figure}

\newpage
~
\newpage

\begin{figure}[t] 
\centering
\includegraphics[width=0.99 \textwidth]{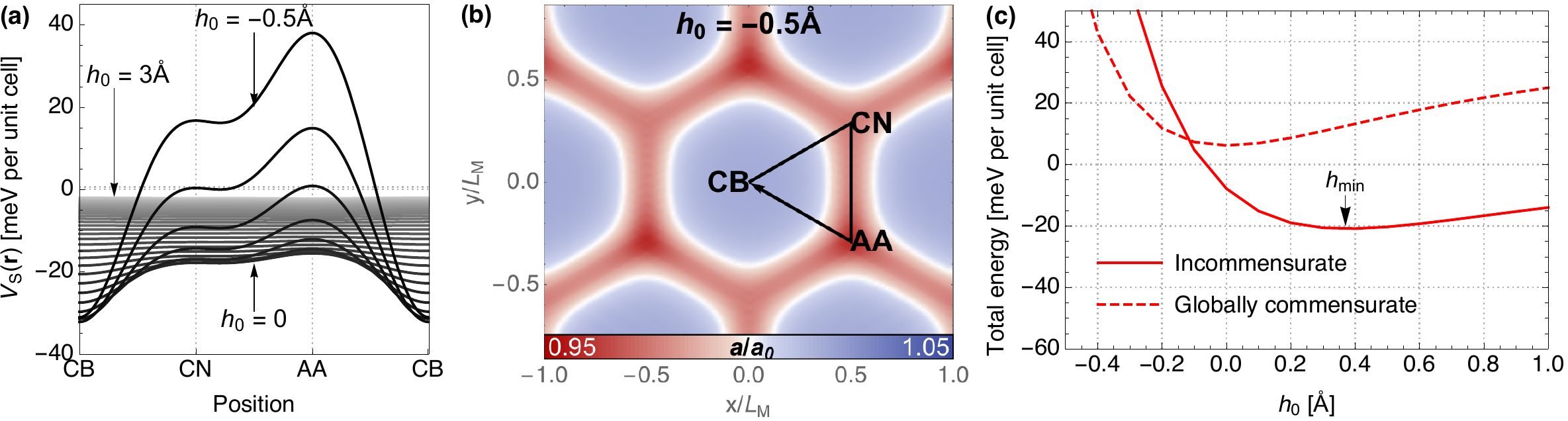} 
\caption{Graphene strain and energy under a flat metal plate. (a) Adhesion energy per graphene unit cell, assuming a pinning of the sample/substrate distance to a constant $z=d_\mathrm{CB}+h_0$ (e.g. by contacting the sample with a flat metallic plate at $z$). Note that for negative $h_0$ adhesion energy differences are enhanced (analogous to high hydrostatic pressure), while a positive $h_0$ suppresses adhesion differences. (b) Equilibrium normalized lattice constant $a/a_0$ profile in the incommensurate phase with $h_0=-0.5$\AA (similar to the $P=40$ GPa case of Supplementary Figure~\ref{fig:pressure}(d)). (c) Total energy relative to graphene in vacuum ($h_0=\infty$). The globally commensurate phase (dashed) becomes energetically favorable for $h_0<-0.12$\AA.}
\label{fig:pressurez}
\end{figure}

\newpage
~
\newpage

\begin{figure}[t]
\centering
\includegraphics[width=0.9 \columnwidth]{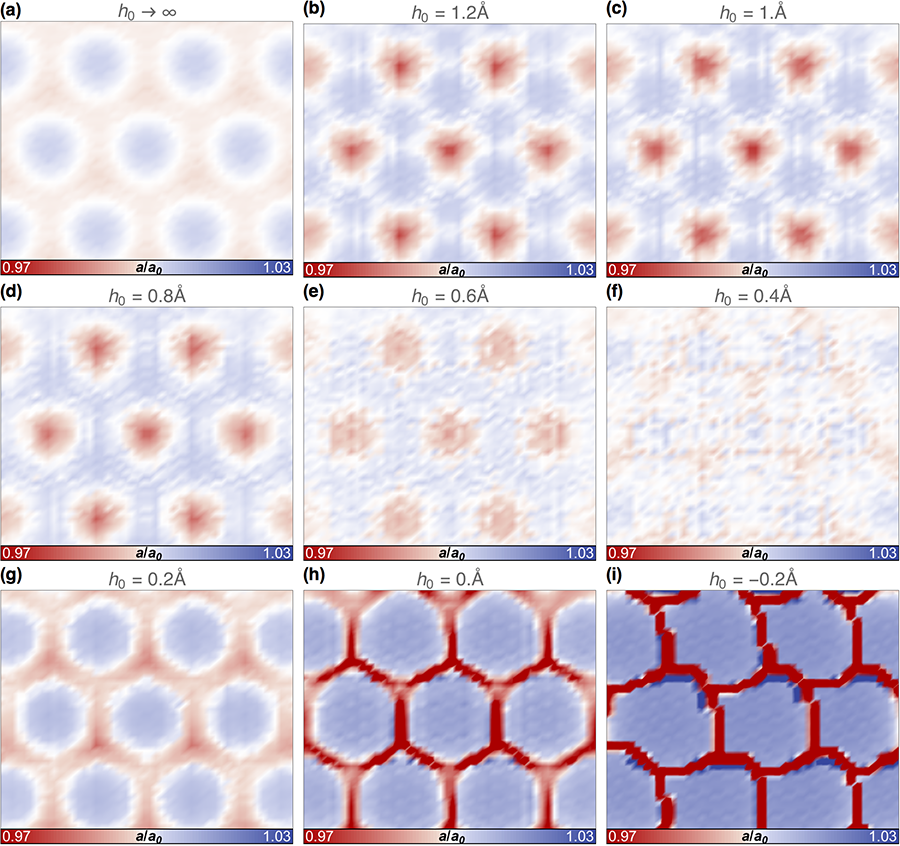}
\caption{Dynamical strain as measured by a tip scanning at different tip-sample distances $h_0$. The tip has a radius $R_\mathrm{tip}=200$ nm, and the maximum lift height is $h_\mathrm{max}=1.5$ \AA. Panel (a) shows the equilibrium (static) strain without the tip ($h_0>h_\mathrm{max}$). Panels (b)-(e) show the attractive regime $h_\mathrm{min}<h_0<h_\mathrm{max}$, where $h_\mathrm{min}=0.37$\AA. Panel (f) is at a crossover $h_0\approx h_\mathrm{min}$. Panels (g) and (h) show the non-hysteretic repulsive regime $0\lesssim h_0<h_\mathrm{min}$, and (i) shows the hysteretic repulsive regime $h_0\lesssim 0$. The small hexagonal mesh in each is an artifact of the mesh size of the simulation. 
}
\label{fig:dynstrain}
\end{figure} 

\newpage
~
\newpage

\begin{figure}[t]
\centering
\includegraphics[width=0.9 \columnwidth]{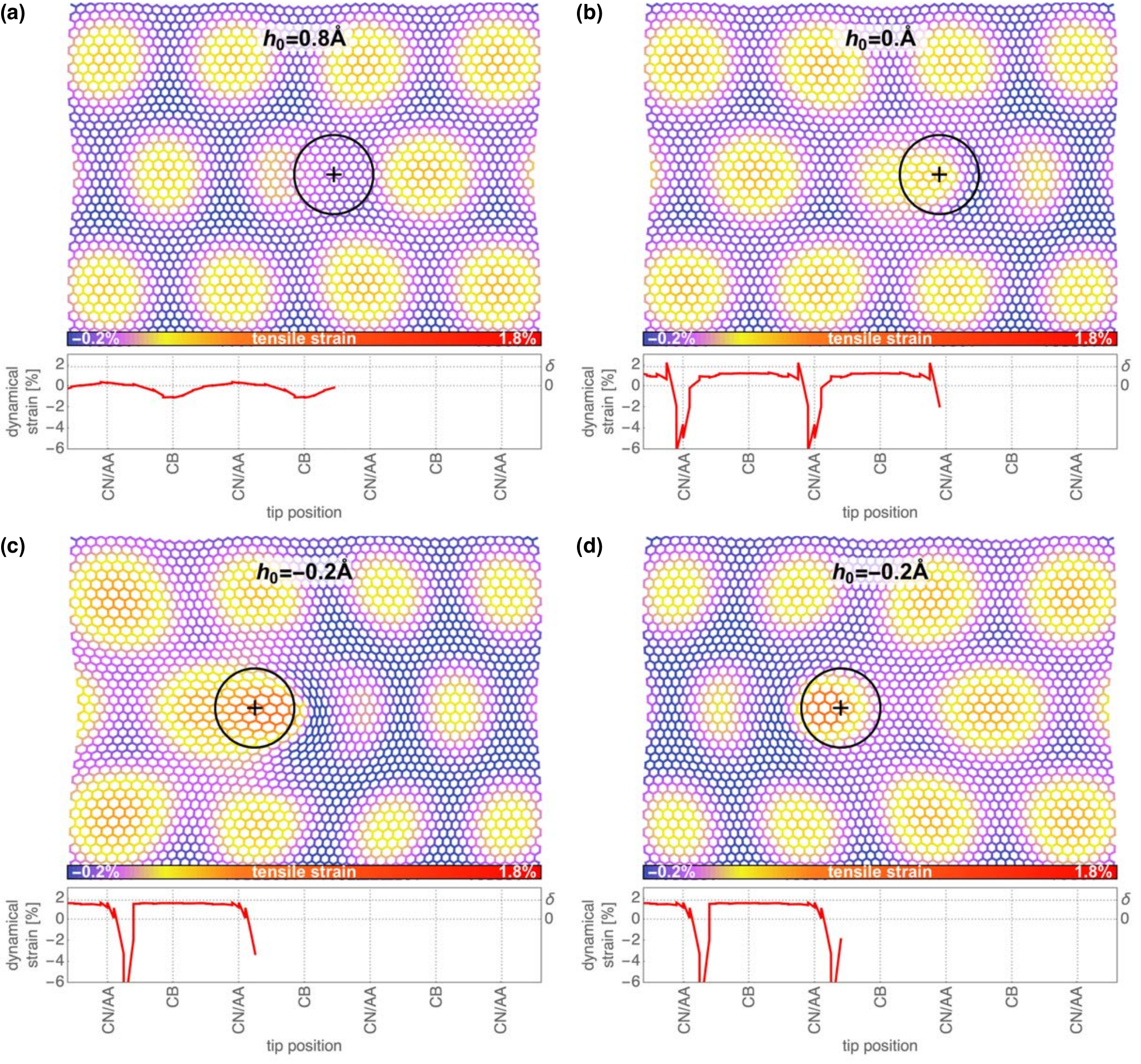}
\caption{Snapshots of Supplementary Movies 1-4. The snapshots depict the time evolution of both the instantaneous strain (top view) and the dynamical strain (bottom view) under a tip, scanning from left to right across the middle of the sample. We show the attractive (a), the repulsive non-hysteretic (b) and the repulsive hysteretic regimes (c) and (d). The tip position in the plane is shown in black, with its distance to the sample shown by the $h_0$ label. Both the lattice constant and the spatial deformation have been scaled up for better visibility in the top views (the color scale and bottom views represent the actual, unscaled strain).}
\label{fig:frames}
\end{figure} 

\newpage
~
\newpage

\begin{figure}[t]
\centering
\includegraphics[width=0.9 \columnwidth]{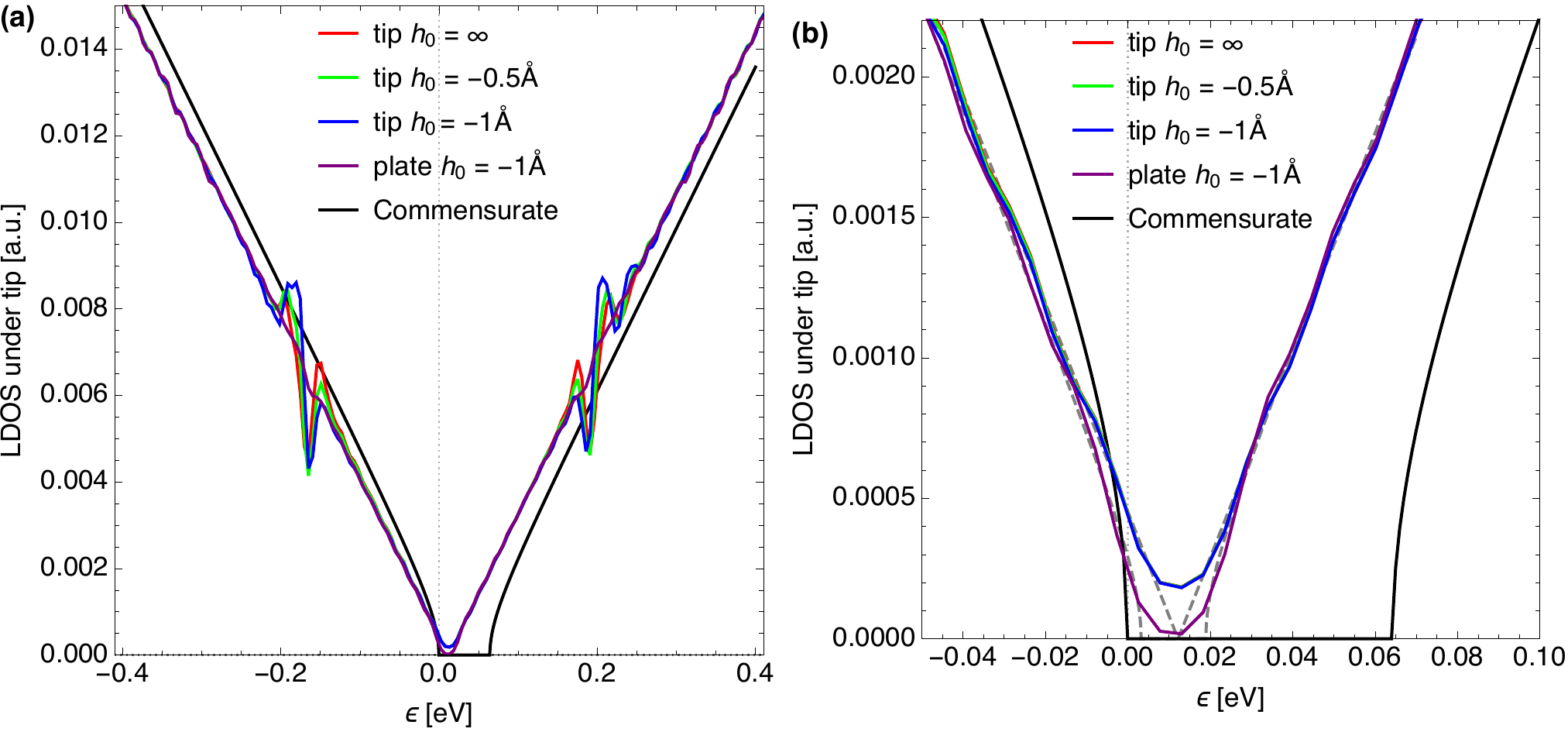}
\caption{Local density of states (LDOS) under the tip (at the center of a CB stacking region) for different tip heights $h_0$, including the effect of pressure-induced deformations. In purple, the LDOS for a flat plate at height $h_0=-1$\AA. The dashed line in the zoom-in (panel b), correspond to LDOS fits to a gapped Dirac spectrum, that show a negligible gap under tip pressure (because of the finite area of the deformed region), and the finite gap under the plate. The black line corresponds to the LDOS for the globally commensurate phase, with a $~\sim 64$ meV gap without many-body corrections.}
\label{fig:ldos}
\end{figure} 

\newpage
~
\newpage

\begin{figure}[t]
\centering
\includegraphics[width=0.5 \columnwidth]{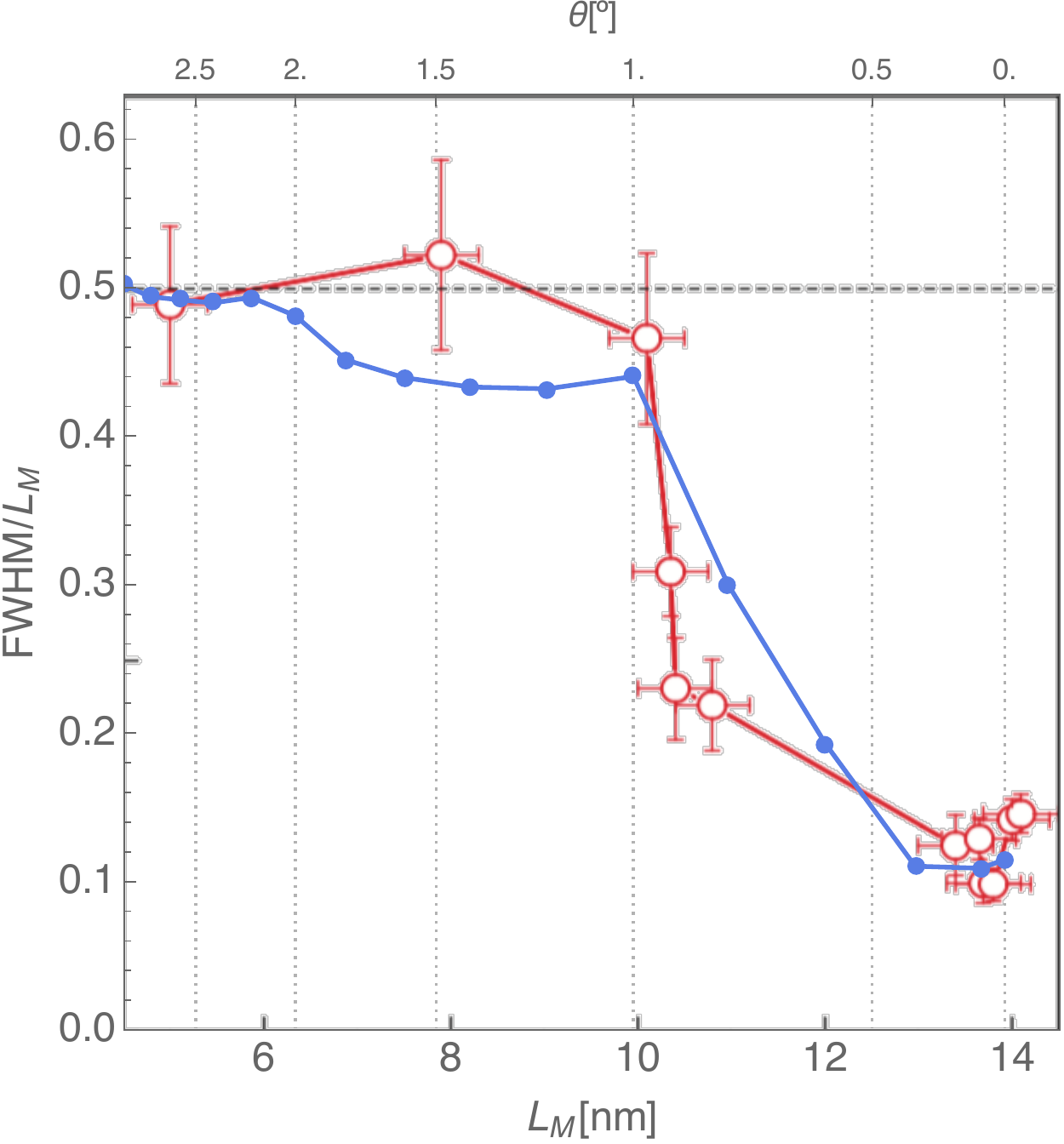}
\caption{Apparent lattice strain versus moir\'e period. In blue, full-width at half-maximum (FWHM) of the apparent strain, ~\ref{dynstrain}, across the center of a stacking boundary, as a function of twist angle $\theta$. In red, FWHM reported in Ref.~\cite{Woods2014}, reproduced here with permission.}
\label{fig:Manchester}
\end{figure} 

\newpage
~
\newpage

\section*{Supplementary Tables}

\begin{table}[!ht]
\centering
\begin{tabular}{ c c | c c | c c } 
  $d_\mathrm{AA} \textrm{[\AA]}$ & $V_\mathrm{AA} \left[\frac{\textrm{meV}}{\textrm{cell}}\right]$ & $d_\mathrm{CB} \textrm{[\AA]}$ & $V_\mathrm{CB} \left[\frac{\textrm{meV}}{\textrm{cell}}\right]$ & $d_\mathrm{CN} \textrm{[\AA]}$ & $V_\mathrm{CN} \left[\frac{\textrm{meV}}{\textrm{cell}}\right]$ \\
  \hline
   $3.67$ & $-15.6$ & $3.11$ & $-32.2$ & $3.55$ & $-17.9$
\end{tabular}
\caption{Fit parameters for data in Supplementary Figure~\ref{fig:fit} using model of ~\ref{Vfit}.}\label{tab:fit}
\end{table}

\newpage

\section*{Supplementary Note 1 - Constant Tunneling Current Strain Maps}

The interaction strength between the STM tip and the graphene/hBN heterostructure can be tuned experimentally by varying the tunneling resistance. The tunneling current can be expressed as 
\begin{equation} \label{tunnelcurrent}
I \approx \frac{4 \pi e}{\hbar} e^{-\Delta z\sqrt{\frac{8m\phi}{\hbar^2}}} \rho_\mathrm{t} \int_{-eV}^{0} \rho_\mathrm{s}(\varepsilon) d\varepsilon ,
\tag{Supplementary Equation 1}
\end{equation}
where $\Delta z$ is the relative separation between the tip and sample, $\phi$ is the tunneling barrier height (essentially the average work function of the tip and sample), $\rho_\mathrm{{s(t)}}(\varepsilon)$ is the density of states of the sample (tip), $m$ is the bare electron mass and $e$ is the charge of the electron. Topography measurements are performed using a feedback loop to maintain a constant tunneling current $I$. Through a simple Ohm's law relationship, $R = V/I$, it is easy to see that lowering the tunneling resistance requires either lowering the bias voltage $V$ or raising the tunneling current $I$. In the latter case, ~\ref{tunnelcurrent} implies that this requires lowering the tip-sample separation $z$, thus increasing the vdW interaction strength between the tip and sample.

Controlling the vdW interaction strength through the sample bias $V$ is more complicated however, as the local density of states of the sample depends on energy. Therefore, in principle, large differences in the LDOS as a function of energy could dominate ~\ref{tunnelcurrent}, perhaps artificially changing the appearance of the topography (rather than the dynamic interaction proposed in the main text). Although this is unlikely to also explain the differences in the strain maps, we conclusively eliminate this effect by acquiring topography maps at a fixed sample bias and varying tunneling current. Figs.~4(a) and (b) of the main text were acquired at the same sample bias with varying tunneling current, to ensure that the switch from the attractive to repulsive regimes could be achieved without any LDOS variation. Similarly, the repulsive regime can be driven hysteretic by varying only the tunneling current as well.

\section*{Supplementary Note 2 - Decay of Tunneling Current}

In Fig.~2(a) of the main text, we track the decay of the tunneling current upon tip retraction for nearly-aligned graphene on hBN. In Supplementary Figure~\ref{fig:tunnelingcurrent}(a) we also track the tunneling current upon reapproaching the tip. The solid black and red curves show the current upon tip retraction for graphene on SiO$_2$ and hBN, respectively. Again, we observed exponential decay for SiO$_2$, and an initial region of parabolic decay for hBN, indicative of the compression of the graphene towards the hBN substrate. The dotted lines show the tunneling current upon reapproach of the tip. The SiO$_2$ curves are nearly identical, while the hBN curves exhibit significant hysteresis, with the tunneling current offset to smaller values of $\Delta z$ upon reapproach.

The large hysteresis suggests that the separation between the graphene and the tip depends on the direction of the tip motion. This provides further evidence that the tip is pushing and pulling the graphene, as there should be no significant source of hysteresis between a tip retraction and approach given an immobile graphene sheet. When the tip is only retracted a short distance such that the tunneling current remains finite (blue), the graphene remains stuck to the tip and consequently no hysteresis is observed. The inset of Supplementary Figure~\ref{fig:tunnelingcurrent}(a) shows a similar measurement on hBN taken with a very small initial tunneling current, such that the tip starts far from the graphene. In this case, the decay is much closer to exponential, with virtually no hysteresis.

Finally, we take similar retraction measurements on misaligned graphene on hBN (Supplementary Figure~\ref{fig:tunnelingcurrent}(b)). In this case, we see very similar behavior to the nearly-aligned samples, where the decay of the tunneling current is first parabolic before becoming exponential, and the decay becomes slower as the initial tip-sample separation is decreased (black to red to blue). This demonstrates that the ability to compress or decompress the graphene relative to the hBN is a property of the two materials independent of their relative rotation.

\section*{Supplementary Note 3 - Local Band Gaps}

Recent transport experiments by multiple groups have demonstrated electronic band gaps in nearly-aligned graphene on hBN heterostructures~\cite{Hunt2013,Chen2013,Woods2014,Gorbachev2014,Wang2015b}. Gap sizes as large as about 50 meV have been observed for devices in perfect alignment, and the gap size decreases as the misalignment is increased, a phenomenon not yet fully understood. The origin of the gaps themselves is still under debate, with proposals suggesting the gap either arises from sublattice symmetry breaking in the graphene~\cite{Giovannetti2007,Slawinska2010}, through many-body interactions~\cite{Song2013,Bokdam2014,Slotman2015}, or through a combination of the two.

It is natural, then, to expect to observe these band gaps in STS measurements of nearly-aligned devices, though to this point no such gaps have been reported~\cite{Yankowitz2014b}. In this study, we are also unable to unambiguously identify large band gaps in our dI/dV spectroscopy measurements. To test for such gaps, we take dI/dV as a function of back gate voltage, shown in Supplementary Figure~\ref{fig:gatemap}. We may first look for signatures of a band gap in individual dI/dV curves, as shown in the inset of Supplementary Figure~\ref{fig:gatemap}, taken at charge neutrality. However, the dI/dV shows a characteristic $V$-shape of an ungapped Dirac cone, rather than a gapped Dirac cone, in which the dI/dV should go to zero and exhibit a flat area in energy roughly equal to the magnitude of the gap. This limits the maximum size of the band gap to roughly the magnitude of our ac excitation bias, around 10 meV.

We may also examine the movement of the Dirac point in sample voltage (or energy) as a function of back gate voltage (or carrier density). In an ungapped Dirac cone, the movement of the Dirac point is expected to disperse as the square root of gate voltage~\cite{Zhang2008}. In a gapped cone, there should either be a jump discontinuity in the position of the Dirac point, or a region of linear movement surrounding the Fermi energy (zero sample voltage), depending on the movement of the Fermi energy through the gap. However, we are never able to unambiguously identify a such a departure form the basic square root of gate voltage dispersion in any of our gate maps. 

As a result, we have no evidence of band gaps anywhere near the size observed in global transport measurements. However, this is not totally unexpected. These devices never exhibit globally commensurate states, where a band gap of order 50 meV is expected simply from the potential difference of the two graphene sublattices~\cite{Giovannetti2007,Slawinska2010}. This instead suggests that while some small component of the band gap may emerge from weak sublattice symmetry breaking due to the small equilibrium graphene strain fields, it is likely enhanced significantly through many-body interactions. Unfortunately, our large metal tip, sitting around 1 nm or less away from the sample surface, likely screens out these many-body interactions, leaving a smaller band gap which we are not able to clearly resolve.

To attempt to circumvent this issue, we have taken gate maps in both extremes of the tip position. By acquiring the gate map with the tip very far away from the surface, we can attempt to minimize this screening. Alternatively, by moving the tip very close to the surface, we can attempt to increase the area of commensurate graphene on hBN (via the mechanism central to this work), and in the limit of the entire sample becoming commensurate, the replica Dirac points should disappear and a large band gap may emerge without the need for significant many-body enhancement. However, in no case are we able to observe a band gap. Supplementary Figure~\ref{fig:gatemap} shows the result for a very close tip, and we still observe the replica Dirac points with no unambiguous band gap. This is also not unexpected, as we expect the tip can only modify the graphene lattice on length scales of about one moir\'e unit cell, which is not sufficient to localize electrons and open a gap (see Supplementary Note 5 for a theoretical analysis). The gate maps at very large tip separations look similar to the small separations, suggesting we are not able to move the tip far enough away to reduce possible screening effects in the incommensurate case. While our measurements are unable to directly address the nature of these band gaps, the lack of their observation points to a significant many-body contribution to their magnitude at equilibrium.

\section*{Supplementary Note 4 - Elastic theory for graphene/\MakeLowercase{h}BN moir\'e superlattices}\label{sec:theory}

\subsection*{Geometric setup}

Unstrained graphene deposited on hBN creates a moir\'e pattern controlled by the  interlayer rotation angle $\theta$ and the lattice mismatch $\delta\approx 1.8 \%$ between the crystals (graphene's lattice constant $a_0\approx0.246$ nm is smaller than hBN's $a_0'\approx 0.250$ nm).

The graphene lattice is generated by the primitive vectors $\vect{a}_{1,2}$, of modulus $a_0$, which we write in a matrix form as $\mat{a}=(\vect{a}_1,\vect{a}_2)$ (the $\vect{a}_{1,2}$ are the \emph{columns} of $\mat{a}$). The center of the graphene unit cells are at $\vect{r}_{\vect{n}}=\mat{a}\vect{n}$. Likewise the unit cells of hBN are centred at $\vect{r'}_{\vect{n}}=\mat{a}'\vect{n}$. The two Bravais bases are related by
\begin{equation}
\mat{a}'=\mat{R}\mat{a},
\tag{Supplementary Equation 2}
\end{equation}
where the rotation and scaling transformation $\mat{R}$ is
\begin{equation}
\mat{R}=(1+\delta)\left(
\begin{array}{cc}
\cos\theta & -\sin\theta \\
\sin\theta & \cos\theta
\end{array}
\right)
\tag{Supplementary Equation 3}
\end{equation}

We assume that $\theta$ and $\delta$ are such that the two lattices are commensurate. This implies that there exists a finite moir\'e superlattice, with primitive vectors $\mat{A}=(\vect{A}_1,\vect{A}_2)$, such that
\begin{equation}
\mat{A}=\mat{a}\mat{N}=\mat{a}'\mat{N}',
\tag{Supplementary Equation 4}
\end{equation}
where $\mat{N}$ and $\mat{N}'$ are \emph{integer} $2\times 2$ matrices. As an example, if $\theta=0$ and $\delta=1/55$, we have $\mat{N}=56\times\mathds{1}$ and $\mat{N}'=55\times \mathds{1}$. The period $L_\mathrm{M}=|\vect{A}_i|$ reads,
\begin{equation}\label{LM}
L_\mathrm{M}=\frac{1+\delta}{\sqrt{2+(2+\delta)\delta-2(1+\delta)\cos\theta}}a_0
\tag{Supplementary Equation 5}
\end{equation}
and is plotted in Supplementary Figure~\ref{fig:LM} as a function of $\theta$.

The conjugate momenta of the two lattices are denoted by $\mat{g}=2\pi\mat{a}^{-1}$ and $\mat{g}'=2\pi\mat{a}'^{-1}$, while the momenta of the superlattice are 
\begin{equation}
\mat{G}=2\pi\mat{A}^{-1}=\mat{N}^{-1}\mat{g}=\mat{N}'^{-1}\mat{g}'.
\tag{Supplementary Equation 6}
\end{equation}
Note that matrices $\mat{g}$, $\mat{g}'$ and $\mat{G}$ have the corresponding conjugate momenta ($\vect{g}_{1,2}$ etc.) as \emph{rows}, not columns.

In general, we can always write $\mat{N}$ and $\mat{N}'$ in terms of $\mat{R}$. To do this we assume that in the unit cell there is a single moir\'e beating (it is a minimal cell). In this case $\mat{G}=\mat{g}-\mat{g}'$. This allows us to write

\begin{align}
\mat{N}&=\mat{a}^{-1}(\mathds{1}-\mat{R}^{-1})^{-1}\mat{a}\tag{Supplementary Equation 7}
\\
\mat{N}'&=\mat{a}^{-1}(\mat{R}-\mathds{1})^{-1}\mat{a} \notag
\end{align}

which implies also
\begin{align}
\label{Neqs}
\mat{N}'^{-1}\mat{N}&=\mat{a}^{-1}\mat{R}\mat{a}=\mat{a}^{-1}\mat{a}'\tag{Supplementary Equation 8}\\
\mat{N}-\mat{N}'&=\mathds{1}\nonumber\\
\mat{N}'\mat{N}^{-1}&=\mat{a}^{-1}\mat{R}^{-1}\mat{a}=\mathds{1}-\mat{N}^{-1}\nonumber
\end{align}
(In all the expressions above, we may change $\mat{a}$ to $\mat{a}'=\mat{R}\mat{a}$ and they still hold.)

\subsection*{Potential created by the \MakeLowercase{h}BN substrate}

The hBN substrate creates a van der Waals potential that attracts the graphene sample. This adhesion, however, depends on the local stacking across the moir\'e, and is strongest for carbon-boron Bernal alignment. Those parts of the moir\'e are thus more strongly bound to the substrate than the other regions. The spatially varying adhesion landscape makes graphene deform elastically with a three-dimensional displacement field $\vect{u}(\vect{r})=(u_x,u_y,h)$. This deformation expands the preferred carbon-on-boron regions, while other regions contract. To describe this elastic-adhesion interplay we first model the adhesion potential.

We will assume that the hBN substrate is a rigid crystal. Microscopic simulations~\cite{Wijk2014} employing realistic carbon potentials have characterised the graphene/hBN adhesion energy per graphene unit cell for different perfect stackings as a function of interlayer distance $z$, i.e. $V_\alpha(z)$, where $\alpha=\mathrm{AA}$ (aligned lattices), $\alpha=\mathrm{CB}$ (carbon-on-boron), and $\alpha=\mathrm{CN}$ (carbon-on-nitrogen) are the three different perfect stackings. The results are reproduced in Supplementary Figure~\ref{fig:fit}. We find that these results can be accurately fitted by the following model
\begin{equation}\label{Vfit}
V_\alpha(z)=-\frac{5}{6}V_\alpha\left[\left(\frac{d_\alpha}{z}\right)^{11}-\frac{11}{5}\left(\frac{d_\alpha}{z}\right)^5\right],
\tag{Supplementary Equation 9}
\end{equation}
where $d_\alpha$ and $V_\alpha$ represent the equilibrium distance and adhesion potential, respectively.
The fit is shown in see Supplementary Figure~\ref{fig:fit}, and the resulting fitting values (in $\textrm{\AA}$ and meV per graphene unit cell) are shown in Supplementary Table~\ref{tab:fit}.

In addition to the adhesion energy $V_\alpha$, the equilibrium interlayer distance $d_\alpha$ varies locally with the stacking configuration. In the absence of external perturbations, the graphene/hBN moir\'e pattern is thus spontaneously corrugated, with an out-of-plane displacement $h(\vect{r})$ relative to the $z=d_\mathrm{CB}$ plane, so that the interlayer distance across the sample
\begin{equation}
z(\vect{r})=d_\mathrm{CB}+h(\vect{r})
\tag{Supplementary Equation 10}
\end{equation}
interpolates between the different $d_\alpha$, see Supplementary Figure~\ref{fig:profiles}(a) and top (light) curve in Supplementary Figure~\ref{fig:pressure}(a).
Here $\vect{r}=(x,y)$ is the position in the plane, with $\vect{r}=0$ chosen at a carbon-boron stacking point. The out-of-plane corrugation has a maximum amplitude of around
\begin{equation}\label{zcorr}
h_\mathrm{corr}=\max h(\vect{r})=d_\mathrm{AA}-d_\mathrm{CB}\approx 56 \;\textrm{pm}. 
\tag{Supplementary Equation 11}
\end{equation}
We observe similar height fluctuations in our STM measurements, and numerous other groups have also observed these in AFM measurements as well~\cite{Dean2013,Yang2013,Tang2013,Woods2014,Gallagher2015}.

The adhesion potential $V_\mathrm{S}(\vect{r})$ is similarly modulated, see Supplementary Figure~\ref{fig:profiles} and top (light) curve in Supplementary Figure~\ref{fig:pressure}(b). The adhesion energy of a graphene unit cell centred at $\vect{r}$ will be given by $V_\mathrm{S}(\vect{r})$, which interpolates between the different adhesion minima $V_\alpha$. In this model we assume the adhesion potential (and also the equilibrium corrugation) can be approximated by its six lowest harmonics (first star), namely $\pm \vect{g}'_{1,2,3}$, where we have defined the extra momentum $\vect{g}'_3=-(\vect{g}'_1+\vect{g}'_2)$. We encode this first star by dimensionless vectors 
\begin{equation}\label{nu}
\begin{array}{lcr}
{\vect{\nu}}_1=(1,0), & {\vect{\nu}}_2=(0,1), & {\vect{\nu}}_3=(-1,-1),
\end{array}
\tag{Supplementary Equation 12}
\end{equation}
so that 
\begin{equation}\label{adhesion}
V_\mathrm{S}(\vect{r})=v_0+\Re\left\{v_1\sum_{i=1}^{3}e^{i {\vect{\nu}}_i\mat{g}'\vect{r}}\right\}
\tag{Supplementary Equation 13}
\end{equation}
for some real $\phi_0$ and complex $\phi$. This specific form of the potential assumes that $V_\mathrm{S}(\vect{r})$ has an extremum at the center and corners $\vect{r}_\alpha$ of the hexagonal hBN unit cell. The potential at the three extrema $V_\alpha=V_\mathrm{S}(\vect{r}_\alpha)$ are encoded into $v_0$ and $v_1$,
\begin{align}
v_0&=(V_\mathrm{AA}+V_\mathrm{CN}+V_\mathrm{CB})/3\nonumber\\
v_1&=-\frac{V_\mathrm{AA}+V_\mathrm{CN}-2V_\mathrm{CB}}{9}-i\frac{V_\mathrm{AA}-V_\mathrm{CN}}{3\sqrt{3}}\label{v1}\tag{Supplementary Equation 14}
\end{align}
The corrugation $h(\vect{r})$ in Supplementary Figure~\ref{fig:profiles}(a) was built using this same procedure, with $d_0$ and $d_1$ defined as $v_{1,2}$ above, albeit with $d_\alpha-d_\mathrm{CB}$ in place of $V_\alpha$,
\begin{equation}\label{corr}
h(\vect{r})=d_0+\Re\left\{d_1\sum_{i=1}^{3}e^{i {\vect{\nu}}_i\mat{g}'\vect{r}}\right\}.
\tag{Supplementary Equation 15}
\end{equation}

Consider now a perfectly flat and unstrained graphene sample at the plane $z=d_\mathrm{CB}$. The unstrained graphene cells are centred at 
\begin{equation}
\vect{r}_{\vect{n}}=\mat{a}\vect{n},
\tag{Supplementary Equation 16}
\end{equation}
with an integer vector $\vect{n}$. We now consider a lattice distortion field 
\begin{equation}
\vect{u}(\vect{r})=(u_x,u_y,h)
\tag{Supplementary Equation 17}
\end{equation}
The unit cells will be displaced to $\vect{R}_{\vect{n}}=\vect{r}_{\vect{n}}+\vect{u}\left(\vect{r}_{\vect{n}}\right)=\mat{a}\vect{n}+\vect{u}_{\vect{n}}$. The total adhesion energy in a supercell may be written as a sum over the set of graphene $\vect{n}$ vectors (a total of $\det\mat{N}$) that span the supercell
\begin{align}\label{US1}
U_\mathrm{S}=\sum_{\vect{n}}^\textrm{supercell}V_\mathrm{S}\left(\vect{R}_{\vect{n}}\right)
=\sum_{\vect{n}}^\textrm{supercell}\left(v_0+\Re \,v_1\sum_{i=1}^{3} e^{-i{\vect{\nu}}_i \mat{G}(\vect{r}_{\vect{n}}-\mat{A}\mat{a}'^{-1} \vect{u}_{\vect{n}})}\right)
\tag{Supplementary Equation 18}
\end{align}
Here we have used $\vect{u}_{\vect{n}}=\mat{a}'^{-1}\mat{a}'\vect{u}_{\vect{n}}$ and $\mat{g}'\mat{a}=(\mat{g}-\mat{G})\mat{a}=2\pi-\mat{G}\mat{a}$, (recall that $\mat{G}=\mat{g}-\mat{g}'$).
We have also used $\mat{g}'\mat{a}'=2\pi$ and the fact that $\exp(i2\pi {\vect{\nu}}_i\vect{n})=1$, since $\vect{\nu}_i$ and $\vect{n}$ are both integer vectors. The expression for $U_\mathrm{S}$ above can be recast into an integral form at small angles, when $\det \mat{N}$ is large, since the terms become smooth in $\vect{n}$. Instead of $\sum_{\vect{n}}^\mathrm{supercell}$ one may do an integral $\frac{1}{\mathrm{det}{A}}\int_\mathrm{supercell} d^2r$. This also allows one to rediscretize the $U_\mathrm{S}$ sum with \emph{any} mesh that covers the supercell, even one that is much coarser than the atomic mesh, for example $\vect{r}_{\vect{n}}=\mat{b}\vect{n}$, with $\mat{b}=\mat{A}/m$, with $m$ a small integer, e.g. $m=4$ or $5$. When rediscretizing, one should be careful to normalize the sum by the Jacobian $\det\mat{b}/\det{\mat{a}}$.

In practice this rediscretization works well because the deformation fields $\vect{u}(\vect{r})$ that result from this model are smooth on the moir\'e lengthscale $L_\mathrm{M}$, so one needs only a few ($m$) points to within one $L_\mathrm{M}$ to accurately describe the deformation. 

\subsection*{Elastic energy}

The elastic energy $U_\mathrm{E}$ per supercell of a graphene deformation $\vect{u}(\vect{r})$ that is smooth on the atomic spacing is given by continuum elasticity theory,
\begin{equation}
U_\mathrm{E}=\frac{1}{2}\int_{\mat{A}} d^2r\left[
2\mu \,\mathrm{Tr}(\mat{u}^2)+\lambda \left(\mathrm{Tr} \,\mat{u}\right)^2\right]  , 
\tag{Supplementary Equation 19}
\end{equation}
Here $\mat u=u_{ij}=\frac{1}{2}(\partial_i u_j+\partial_j u_i+\partial_i h\partial_j h)$ is the strain, and $\lambda\approx 3.5~ \mathrm{eV/\AA^2}$ and $\mu\approx 7.8~\mathrm{eV/\AA^2}$ are the Lam\'e factors for graphene. In the following, the quartic in $h$ contribution to $U_\mathrm{E}$ will be neglected, since it is of the order of $(2h_\mathrm{corr}/L_\mathrm{M})^2\approx 10^{-5}$. This approximation decouples the equilibrium corrugation from in-plane strains, so that $h(\vect{r})$ is always given by ~\ref{corr}.

To evaluate $U_\mathrm{E}$, one needs to approximate the derivatives $\partial_iu_j$ in $\mat{u}$ by finite differences in the two dimensional mesh $\vect{r}_{\vect{n}}=\mat{b}\vect{n}$. Since this mesh is triangular in this case, the finite differences are best evaluated at the center of each triangle, i.e. in the dual honeycomb lattice formed by all the triangle centers. When thus evaluating the integral as a discrete sum of finite differences, care must be taken once more to properly normalize to the total supercell area divided by the number of dual mesh points (two) per mesh unit cell
\begin{equation}\label{UE}
U_\mathrm{E}=\frac{\det\mat{b}}{2}\sum^\textrm{dual mesh}_{\vect{n}} \frac{1}{2}\left[
2\mu \,\mathrm{Tr}(\mat{u}_{\vect{n}}^2)+\lambda \left(\mathrm{Tr} \,\mat{u}_{\vect{n}}\right)^2\right] 
\tag{Supplementary Equation 20}
\end{equation}

\subsection*{Equilibrium strains under uniform pressure}

The (tensile) strain profile of the graphene sample at equilibrium is defined by the field
\begin{equation}
\frac{1}{2}\mathrm{Tr}\mat{u}(\vect{r})=\frac{a}{a_0}-1
\tag{Supplementary Equation 21}
\end{equation}
where $a$ is the average lattice constant of the sample at point $\vect{r}$. This strain profile arises in the sample at equilibrium as a result of the forces derived from the total elastic plus adhesion potential $U=U_\mathrm{E}+U_\mathrm{S}$. The expected strain using our model for $U$ is shown in Supplementary Figure~\ref{fig:pressure}(c), with a cut along the white line shown in Supplementary Figure~\ref{fig:pressure}(e), top (light) curve. This strain is computed by minimizing $U(\vect{u}_{\vect{n}})$ as a function of the discretized in-plane disortion field $\vect{u}(\vect{r}_{\vect{n}})=\vect{u}_{\vect{n}}$, while $h(\vect{r})$ is fixed by the equilibrium value, ~\ref{corr}, see Supplementary Figure~\ref{fig:profiles}(a) and Supplementary Figure~\ref{fig:pressure}(a), top (light) curve. The numerical optimization of $U$ over in plane distortions $\vect{u}_{\vect{n}}$ is efficiently implemented using the conjugate gradient method. The fact that the continuum approximation, ~\ref{US1}, of the adhesion potential allows for a discretization mesh of the disortion field $\vect{u}_{\vect{n}}$ that is conveniently coarser than the atomic spacing, allows for important numerical efficiency gains. 

We see in Supplementary Figure~\ref{fig:pressure}(c) that CB-stacked regions are slightly expanded relative to graphene in vacuum, while AA, and to less extent also CN, are correspondingly compressed. Since this minimization assumes no global expansion of the sample in response to the adhesion, the spatial integral of $\frac{1}{2}\mathrm{Tr}\mat{u}(\vect{r})$ is zero. The variation of strain is however rather small, around $-0.2\%$ at AA, $+0.3\%$ at CB. 

The magnitude of this modulation is controlled by the adhesion energy of favoured CB regions relative to unfavourable AA regions, see top (light) curve of Supplementary Figure~\ref{fig:pressure}(b). It is reasonable to expect that enhancing this adhesion difference one could also enhance the spontaneous strain modulations. This is in principle simple to do. If we apply uniform hydrostatic pressure to the sample, graphene is pushed towards the substrate, which should enhance the adhesion difference, given the adhesion curves $V_\alpha(z)$ in Supplementary Figure~\ref{fig:fit}. To confirm this, we add one more term $V_\mathrm{P}(z)=  P (z-d_\mathrm{CB}) \det \mat{a}$ (the chosen $z$ origin is arbitrary) to the substrate adhesion potentials $V_\alpha(z)$ in ~\ref{Vfit}, where $P$ is pressure. One then performs the minimization of $V_\alpha(z)+V_\mathrm{P}(z)$, to find the modified values of $V_\alpha$ and $d_\alpha$ as a function of pressure $P$. As expected, the $d_\alpha$ decrease and $|V_\mathrm{CB}-V_\mathrm{AA}|$ is strongly enhanced. This is shown in Supplementary Figure~\ref{fig:pressure}(a) and (b) in the range $P=0$ to $P=40$ GPa. The resulting equilibrium strain under pressure is shown in Supplementary Figure~\ref{fig:pressure}(d) and (e). They show a positive pressure-induced expansion at $CB$ stacking regions that saturate for high pressures at the commensurate limit $a/a_0=1+\delta$, i.e. a $1.8\%$ tensile strain so that graphene becomes locally commensurate to the underlying hBN crystal. The boundaries between CB regions become narrow and strongly compressed, reaching a tensile strain of $\sim -4\%$ at $P=40$ GPa. The total energy per supercell of the sample 

In all the above simulations the area of the sample is kept constant as pressure is increased. In particular, we did not allow up to now for the possibility of the sample developing a global uniform expansion $a/a_0=1+\delta$ to conform to the substrate everywhere. While this configuration is indeed not energetically favorable under zero pressure, one can expect that at high-enough pressures, the enhanced adhesion differences that lead to expanded $CB$ regions discussed above would also favor a globally commensurate phase. To evaluate the possibility of a structural transition into said phase, we compute its energy as a function of pressure, relative to that of  graphene in vacuum ($U=0$). The energy per unit cell for the globally commensurate phase reads
\begin{equation}
V_\mathrm{CB}(P)+2\det\mat{a}(\lambda+\mu)\delta^2 
\tag{Supplementary Equation 22}
\end{equation}
and is shown as a function of pressure by the dashed line in Supplementary Figure~\ref{fig:pressure}(f). It indeed becomes smaller than the energy of the sample with the moir\'e strain profile (solid line) for pressure above a critical value $P_\mathrm{c}=0.15 GPa$. We thus expect that as soon as pressure exceeds $P_\mathrm{c}$, the graphene sample would undergo a (first-order) structural transition into global commensuration. Electronically, this phase is expected to develop a large gap at the Dirac point around $50-200$ meV~\cite{Giovannetti2007,Bokdam2014,SanJose2014b}.

An alternative method to hydrostatic pressure to enhance moir\'e strains is to apply pressure with a metallic plate. Graphene adheres to most metals more strongly than to hBN, so that a perfectly flat metallic plate at constant $z(\vect{r})=d_\mathrm{CB}+h_0$ pressing onto the sample would completely suppress sample corrugations, so that $h(\vect{r})\approx h_0$ (we have incorporated the equilibrium plate-graphene distance into $h_0$ here). As a result, $V_\alpha$ would not correspond to the minimum of $V_\alpha(z)$ but rather to $V_\alpha(d_\mathrm{CB}+h_0)$ at constant the $z$. Supplementary Figure~\ref{fig:pressurez}(a) shows the corresponding $V_\mathrm{S}(\vect{r})$ for different values of $h_0$ along a $CB-CN-AA-CB$ spatial path. The equilibrium strains for $h_0=-0.5$\AA \space are shown in panel (b). Note the expanded CB regions, and the narrow boundaries, analogous to the case of $P=40$ GPa hydrostatic pressure, with a similar range of tensile strains from $1.8\%$ (CB) to $\sim -4\%$ (AA). Panel (c) shows the total energy as a function of $h_0$, both for the incommensurate and the globally commensurate phase. All energies are taken relative to that of graphene in vacuum ($h_0\to \infty$, i.e. $U=0$). We see that, as long as the constraint $z_0=d_\mathrm{CB}+h_0$ is uniform across the sample, graphene is expected to spontaneously snap into a globally commensurate phase for $h_0\lesssim -0.12$\AA. As for the case of hydrostatic pressure, the transition is first order, and is expected to be thermally activated.

It is also interesting to note that the total energy of the incommensurate phase has a minimum at a value of $h_0$ around 
\begin{equation}\label{hmin}
h_\mathrm{min}=0.37~\textrm{\AA},
\tag{Supplementary Equation 23}
\end{equation}
somewhat smaller than the corrugation $h_\mathrm{corr}$ in the equilibrium sample, which implies that within the commensurate phase, a repulsive (attractive) force will develop below (above) this position between graphene and the metallic plate. 


\subsection*{Tip potential}

The effect of a large metallic tip close to the sample is similar to the above analysis of pressure by a metallic plate. In the case of a metallic tip, $h_0$ should be taken to be position dependent. For a paraboloid-like tip of radius $R_\mathrm{tip}$ with its apex at $(x_0,y_0,h_0)$, we have
\begin{equation}
h_\mathrm{tip}(\vect{r})=h_0+\frac{(x-x_0)^2+(y-y_0)^2}{2R_\mathrm{tip}}
\tag{Supplementary Equation 24}
\end{equation}

Typical tip radii, around $R_\mathrm{tip}\approx 200$ nm, are quite large compared to the moir\'e lengthscale $L_\mathrm{M}\lesssim 14$ nm, which justifies the above paraboloid model. Graphene does not conform to the tip at all positions, however, since it is constrained by boundary conditions to remain stuck to hBN far from the tip (this precludes a tip-induced transition into a globally commensurate phase). The tip constraint should therefore be truncated to values of $h$ below a maximum retraction value $h_\mathrm{max}$, so that
\begin{align}\label{htip}
h(\vect{r})=\left\{\begin{array}{l}
h_\mathrm{tip}(\vect{r})\textrm{ if } h_\mathrm{tip}(\vect{r})<h_\mathrm{max}\\
h_\mathrm{eq}(\vect{r})\textrm{ otherwise}
\tag{Supplementary Equation 25}
\end{array}\right.
\end{align}
This profile is illustrated in Fig.~1(c) of the main text. The value of $h_\mathrm{max}$ can be estimated to be $h_\mathrm{max}\approx 1-2$ \AA\space from the crossover regime in Fig.~2(b) of the main text. The interaction range of the tip, i.e. the distance $R_\mathrm{max}$ from its apex below which the sample/substrate adhesion is controlled by the tip, reads, for tip height $h_0<h_\mathrm{max}$,
\begin{equation}\label{tiprange}
R_\mathrm{max}=\sqrt{2R_\mathrm{tip}(h_\mathrm{max}-h_0)}
\tag{Supplementary Equation 26}
\end{equation}
If $h_0>h_\mathrm{max}$, the sample is not adhered to the tip, and we assume $R_\mathrm{max}=0$.

As revealed by the tip retraction experiment of the main text, the sample $h(\vect{r})$ does not exactly conform to the tip $h_\mathrm{tip}(\vect{r})$, and it furthermore recedes from the tip less abruptly as $h_0$ exceeds $h_\mathrm{max}$, but the above model proves to be rather accurate to describe the mechanical tip-sample interactions.

The influence of the tip on the sample strain throughout a scan can be incorporated rather economically into the adhesion potential by spatially modulating the value of $v_0$ $v_1$,
\begin{align}
U_\mathrm{S}=\frac{\det\mat{b}}{\det\mat{a}}\sum_{\vect{n}}^\textrm{supercell}\left(v_0(\vect{r}_{\vect{n}})+\Re \,v_1(\vect{r}_{\vect{n}})\sum_{i=1}^{3} e^{-i{\vect{\nu}}_i \mat{G}(\vect{r}_{\vect{n}}-\mat{A}\mat{a}'^{-1} \vect{u}_{\vect{n}})}\right)\label{US}
\tag{Supplementary Equation 27}
\end{align}
where $v_0(\vect{r}), v_1(\vect{r})$ are defined as in ~\ref{v1}, but with $V_\alpha(d_\mathrm{CB}+h(\vect{r}))$ from \ref{Vfit} and \ref{htip} in place of constants $V_\alpha$.

\subsection*{Scanning tip and dynamical strain}

A modulation of the graphene/hBN distance $z(\vect{r})=d_\mathrm{CB}+h(\vect{r})$ imposed by an STM tip produces a spatial variation in the total energy density accumulated in the deformation field. The total energy stored in the deformation field has a minimum at $h_\mathrm{min}=0.37$\AA. A gradient in $h(\vect{r})$ will therefore give rise to forces that will tend to push the moir\'e strain profile, and in particular the boundaries between locally commensurate CB regions, towards points with $h(\vect{r})=h_\mathrm{min}$. Effectively, therefore, a tip at a height $h_0<h_\mathrm{min}$ will tend to push stacking boundaries away from its apex, while if $h_0>h_\mathrm{corr}$ the tip will appear to attract stacking boundaries. These forces dynamically modify the deformation field as the STM tip scans the sample. In this section we analyze this dynamical response, and its consequences for the apparent strain measured by the tip.

In the preceding section, the total energy per supercell $U=U_\mathrm{E}+U_\mathrm{S}$ in the presence of a tip has been expressed as a function of distortions $\vect{u}_{\vect{n}}$ and rotation angle $\theta$ on a conveniently coarse discretization of the moir\'e pattern. All parameters of the model are known to reasonable precision, including the tip radius $\vect{R}_\mathrm{tip}$ and the maximum retraction distance $h_\mathrm{max}$. It is then possible to minimize the total energy numerically to obtain the equilibrium elastic configuration for each tip position as it scans the sample. As in the cases of hydrostatic and plate pressure, this is done using conjugate gradient methods. The result is a discretisation of the equilibrium deformation $\vect{u}(\vect{r}, \vect{r}_0, h_0)$, and the associated strain tensor $\mat{u}(\vect{r}, \vect{r}_0, h_0)$, as a function of tip position $\vect{r}_0=(x_0,y_0,h_0)$. 

When relaxing the lattice in response to a scanning tip, it is important to choose as seed to the conjugate gradient method the relaxed configuration from the prior tip position. This choice is irrelevant for larger values of $h_0$, since there is no scanning hysteresis. For smaller $h_0$, however, the tip hysteretically drags the stacking domains along, so the choice of seed is important, as opposite scanning directions yield different configuration paths.

Due to the $\vect{r}_0$ dependence of $\mat{u}$, the graphene tensile strain $a/a_0-1$ (normalized change in the average lattice constant) as measured by the tip (what we dub here `dynamical strain') is not simply the static expansion $\frac{1}{2}\mathrm{Tr}\mat{u}=\frac{1}{2}\left(\partial_xu_{x}+\partial_yu_{y}\right)$. It also acquires a dynamical contribution. The dynamical strain reads
\begin{equation}\label{dynstrain}
\frac{a}{a_0}-1=\frac{1}{2}\left.\left(\partial_xu_{x}+\partial_yu_{y}+\partial_{x_0}u_{x}+\partial_{y_0}u_{y}\right)\right|_{\vect{r}_0=\vect{r}}
\tag{Supplementary Equation 28}
\end{equation}

We have simulated this dynamical strain for a $\theta=0^\circ$ sample, scanned with an STM tip of realistic radius $R_\mathrm{tip}=200$ nm, and a $h_\mathrm{max}=1.5$ \AA. The results for varying tip-sample scanning distances $h_0$ are shown in ~\ref{fig:dynstrain}. Panel (a) shows the static expansion, corresponding to a tip with negligible interaction with the sample, $h_0\to \infty$ (actually $h_0>h_\mathrm{max}$ in our simplified adhesion model). As in Supplementary Figure~\ref{fig:pressure}(c), it has smooth strain profiles in the $\sim -0.2\%$ to $0.3\%$ range, with CB regions expanded relative to the rest. The dynamical strain as the tip scans at $h_0<h_\mathrm{max}$ shows three distinct regimes, which we describe below.

The `attractive regime', panels (b)-(e), corresponds to $h_\mathrm{min}<h_0<h_\mathrm{max}$, see ~\ref{hmin}. In this scanning range the tip locally lifts the sample away from the substrate, irrespective of its position $\vect{r}_0$. Consequently, adhesion differences $|V_\mathrm{CB}-V_\mathrm{AA}|$ are reduced under the tip, but remain the same away from the tip. This produces an effective attraction between the tip and the surrounding the stacking boundaries. As a result, the boundaries are partially dragged along by the tip as it scans, and therefore appear to be expanded (positive dynamical strain). The CB regions, in contrast, exhibit negative dynamical strain (they appear compressed). The latter is a consequence of a basic property of the dynamical strain. Just like the static strain of an asymptotically relaxed sample, the dynamical strain  integrates to zero across the sample, so that a positive dynamical strain of stacking boundaries should be compensated by a negative dynamical strain elsewhere. (This is satisfied as long as the sample as a whole doesn't slide in response to the scanning tip, and that the dynamical strain is not discontinuous, i.e it is non-hysteretic).

A crossover pattern is obtained  at $h_0\approx h_\mathrm{min}$, panel (f), before entering a `repulsive regime' for $h_0<h_\mathrm{min}$. In this scanning range, the tip pushes stacking boundaries away as it moves. As the sample as a whole is assumed to not slide as a result of scanning, the boundaries quickly slide back under the tip when they are pushed beyond a maximum distance. This quick sliding makes them appear compressed far above their equilibrium compression $\sim-0.2\%$. The resulting dynamical strain is shown in panels (g)-(i). If $h_0\gtrsim 0$ this pushing is non-hysteretic (panels (g) and (h)), so that opposite scanning directions follow the same path in the sample configuration space. The dynamical strain maps preserve all the moir\'e symmetries. If $h_0$ is decreased below zero, however, (the tip is pushing the sample strongly enough into the substrate), the area of commensurate CB regions grow further, their static strain approaches the maximum static value $a/a_0-1=\delta$ as in Supplementary Figure~\ref{fig:pressurez}(b) (the dynamical strain is not bounded), and the stacking boundaries are vigorously pushed away from the tip. When boundaries are pushed far enough in this regime, they snap back irreversibly under the tip~\cite{Yankowitz2014}, which gives a discontinuous jump in the dynamical strain. This is a `hysteretic repulsive regime'. The corresponding dynamical strain map (panel (i)) then depends on the scanning direction and breaks the $C_3$ symmetry of the moir\'e.

All these regimes are observed in our experiments, and closely match the simulations above, with one exception. Deep in the hysteretic repulsive regime, the observed dynamical strain may develop one further transition not observed in the simulations, whereby the dynamically compressed stacking boundaries suddenly switch to a positive dynamical strain, well above the maximum static value $\delta=1.8 \%$. This can be understood as the result of reversible sample delamination in front of the tip. The compression accumulated in a pushed boundary can exceed a value where the sample becomes unstable to buckling out of plane in front of the tip. This structural transition greatly relaxes the accumulated compression, which makes the stacking boundary recede further away from the tip, and thus appear to develop a local expansion. The possibility of delamination is not included in our simulations, however, so we can only argue about it on a qualitative level.

Supplementary Movies 1-4 provide animations of the dynamical and instantaneous strain under the scanning tip in the different response regimes. Representative snapshots of these simulations are shown in Supplementary Figure~\ref{fig:frames}.

\section*{Supplementary Note 5 - Local density of states under the tip}\label{LDOS}

The pressure-induced commensuration between graphene and the hBN substrate has powerful implications for the electronic structure. Assuming a hopping amplitude $t_\perp\approx 0.3$ eV between carbon and boron atoms, and a hBN valence band at $\epsilon_\mathrm{B}\approx 1.4$ eV with respect to graphene's Dirac point, the graphene regions with CB-stacking will acquire a substrate induced self-energy that is different in the two sublatices, $\Sigma_\mathrm{A}\approx t_\perp^2/\epsilon_\mathrm{B}$, $\Sigma_\mathrm{B}=0$. This sublattice imbalance $\Delta=\Sigma_\mathrm{A}-\Sigma_{B}\approx 64$ meV takes the form of a mass term $\frac{1}{2}\Delta\sigma_3$ in graphene's Dirac spectrum (plus an unimportant scalar term of equal magnitude). A spatially uniform mass term is expected to open a gap $\Delta$ in graphene's Dirac spectrum. In our samples, however, the pressure-induced CB commensuration is confined to within one moir\'e period approximately, $\lambda_\mathrm{CB}\approx L_\mathrm{M}\approx 14$nm, according to our simulations (see red area in Fig. 1d of main text). A non-straightforward question is whether this area is enough to induce a gap in the local density of states (LDOS) that could be measured by the tip.

A qualitative argument can be used to answer this question. A gap of magnitude $\Delta$ corresponds to the localization of states of typical wavelength $\lambda_\Delta=h v_\mathrm{F}/\frac{\Delta}{2}\approx 165$ nm. Therefore, the minimum spatial extension required of the pressure-induced CB stacking should be of the same order, $\lambda_\mathrm{CB}\gtrsim\lambda_\Delta$, hence much larger than the actual area affected by our tips. 

We have also performed quantitative simulations of the LDOS under the tip, with the graphene lattice subjected to the tip-induced strains obtained with our elastic model.  The local registry between graphene and the hBN substrate create a sublattice- and position-dependent self-energy, while in-plane strains also induce pseudogauge fields \cite{SanJose2014a,SanJose2014b}. The LDOS calculation is performed using the Kernel Polynomial method \cite{Weisse2006}, using the Jackson kernel, and a polynomial order $N=6000$ ($N$ determines the energy resolution of the method). Note that no many-body effect \cite{Song2013,Bokdam2014,Slotman2015} are included here. The computed LDOS is shown in Supplementary Figure~\ref{fig:ldos} for three different values of the the tip height $h_0=-1.0~\textrm{\AA}, -0.5~\textrm{\AA}$ and $\infty$. For comparison we also include the result with a flat plate at height $h_0=-1.0~\textrm{\AA}$, like the one described in Supplementary Figure~\ref{fig:pressurez}, which results in a periodic mesh of thin solitons throughout the sample when artificially precluding a globally commensurate transition. Dashed lines in the zoom (panel b) correspond to fits to a gapped Dirac spectrum of the form $\mathrm{LDOS}(\epsilon)\propto \mathrm{Re}\sqrt{(\tilde\Delta/2)^2-(\epsilon-\epsilon_0)^2}$, where $\tilde\Delta$ is an estimate of the gap that would be obtained for polynomial expansion order $N\to\infty$. Finally, we also include the results of a fully commensurate graphene sheet.

We see that, regardless of the height of the tip, the LDOS around neutrality is the same, and the LDOS gap $\tilde\Delta$ derived from the fit is zero to within less than a millielectronvolt. This is consistent with the qualitative argument above, and with our experimental observations, which do not resolve an LDOS gap regardless of tip pressure. For the plate, however, which induces CB stacking throughout a significant fraction of the sample area, we numerically obtain a finite LDOS gap $\tilde\Delta\approx 16$meV. The ideal value $\tilde \Delta=\Delta$ would be achieved only in the globally commensurate phase (black curves in Supplementary Figure~\ref{fig:ldos}), which is expected to be the true ground state for a plate with $h_0<-0.12$~\AA, as shown in Supplementary Figure~\ref{fig:pressurez}(c).

\section*{Supplementary Note 6 - Comparison to Results of Woods \MakeLowercase{\textit{et al.}}}\label{Manchester}

In a recent work, Woods \textit{et al.} analyzed the morphology, as measured with an AFM tip, of graphene/hBN heterostructures as a function of relative twist angle $\theta$ between the two crystals~\cite{Woods2014}. The main observation was a change in the characteristic width (FWHM) of boundaries between carbon-boron stacking domains. At angles above $1^\circ$, said width appeared to be equal to half a moir\'e period $L_\mathrm{M}(\theta)/2$, with $L_\mathrm{M}$ given in ~\ref{LM}. This is the defining feature of a floating phase, where substrate-induced deformations of graphene are negligible. Below $\theta\approx 1^\circ$, however, the observed width seemed to abruptly transition to a small fraction of $L_\mathrm{M}$, around FWHM$=0.2 L_\mathrm{M}$, as would correspond to strong deformations of graphene due to the substrate. This abrupt jump was attributed to an incommensurate-to-commensurate structural transition, in the language of the Frenkel-Kontorova model~\cite{Chaikin2000}.

Our own observation of the apparent topography and strain of small-angle samples, as measured by an STM tip at small tip separations, closely resembles the observations of Ref.~\cite{Woods2014} for their $\theta<1^\circ$ samples marked by sharp moir\'e walls. Our interpretation, however, is very different from the one proposed by Woods \textit{et al}. Simply by moving the tip away from the sample and repeating the topography measurement, we observe a transition to a smooth moir\'e profile. This clearly indicates that, at least in our samples, the Woods \textit{et al.} interpretation of sharp sample boundaries produced by an equilibrium structural transition of the sample is not valid. The change in apparent strain maps with tip separation demonstrates that our tip is strongly invasive, and is itself the cause of the observed strain profiles of the sample. This is supported by our theoretical modelling, moreover, which suggests that  the sample, when not affected by the tip, should not experience a detectable commensurate-incommensurate transition as a function of angle, and is instead almost perfectly smooth ($\textrm{FWHM}\approx 0.5 L_\mathrm{M}$, floating phase) for angles down to $\theta=0^\circ$. 

To assess whether the apparent transition observed Woods \textit{et al.} could be explained by the influence of the scanning tip as in our experiment, we have simulated, using the theory of Sect.~\ref{sec:theory}, the FWHM of stacking boundaries as a function of angle $\theta=0^\circ \dots 3^\circ$ as measured by a metallic tip. The tip is assumed to scan at $h_0$ = 0.1 \AA~from the sample at all angles, so that at $\theta=0^\circ$ it is in the repulsive, although non-hysteretic regime, with narrow boundaries and a FWHM around $0.1 L_\mathrm{M}$. A sharp increase of the FWHM of stacking boundaries is indeed observed as the twist angle is increased above a certain threshold (just under $1^\circ$ for this tip separation). The change in FWHM across the angle threshold is quite fast, but it is not discontinuous. The precise value of the threshold depends on geometric tip parameters $R_\mathrm{tip}$ and $h_\mathrm{max}$, and corresponds to a moir\'e period approximately equal to the interaction range of the tip (\ref{tiprange}), $R_\mathrm{max}\sim L_\mathrm{M}(\theta)$. Slightly adjusting the tip parameter $h_\mathrm{max}$ to 2.5 \AA~to tune this condition to $\theta\approx 1^\circ$ as in Woods \textit{et al.}, we obtain results compatible with their observations (Supplementary Figure~\ref{fig:Manchester}). This simulation suggests that our interpretation that sharp stacking boundaries at small angles are a tip-induced artifact, not an equilibrium property of the system, could also be relevant for the other scanning probe microscopy measurements of these systems. 

\clearpage

\end{document}